\documentclass[manuscript]{acmart}

\AtBeginDocument{%
  }

\setcopyright{acmlicensed}
\copyrightyear{2018}
\acmYear{2018}
\acmDOI{XXXXXXX.XXXXXXX}

\acmJournal{JACM}
\acmVolume{37}
\acmNumber{4}
\acmArticle{111}
\acmMonth{8}

\usepackage{multirow}
\newcommand{\br}[0]{\textrm{BR}}
\DeclareMathOperator*{\argmax}{arg\,max}

\definecolor{chromeyellow}{rgb}{0.86, 0.52, 0}
\definecolor{applegreen}{rgb}{0.3, 0.6, 0.0}

\newcommand{\cY}{{\color{applegreen}$\checkmark$}}
\newcommand{\cN}{{\color{red}$\times$}}
\newcommand{\cE}{{\color{chromeyellow}$\sim$}}

\newcommand*\rot{\rotatebox[x=0.75cm]{90}}
\newcommand*\rotH{\rotatebox[x=1.25cm]{90}}
\newcommand*\Rot{\rotatebox[x=0.75cm]{80}}

\usepackage{pdflscape}
\usepackage{longtable}

\begin{document}

\title{Game-Theoretic Cybersecurity: the Good, the Bad and the Ugly}

\author{Brandon Collins}
\email{bcollin3@uccs.edu}
\orcid{0000-0002-0440-1026}
\author{Shouhuai Xu}
\email{sxu@ucc.edu}
\author{Philip N. Brown}
\email{pbrown2@uccs.edu}
\affiliation{%
  \institution{University of Colorado Colorado Springs}
  \city{Colorado Springs}
  \state{Colorado}
  \country{USA}
}








\renewcommand{\shortauthors}{Collins et al.}

\begin{abstract}
Given the scale of consequences attributable to cyber attacks, the field of cybersecurity has long outgrown ad-hoc decision-making.
A popular choice to provide disciplined decision-making in cybersecurity is Game Theory, which seeks to mathematically understand strategic interaction.
In practice though, game-theoretic approaches are scarcely utilized (to our knowledge), highlighting the need to understand the deficit between the existing state-of-the-art and the needs of cybersecurity practitioners.
Therefore, we develop a framework to characterize the function and assumptions of existing works as applied to cybersecurity and leverage it to characterize 80 unique technical papers.
Then, we leverage this information to analyze the capabilities of the proposed models in comparison to the application-specific needs they are meant to serve, as well as the practicality of implementing the proposed solution.
Our main finding is that Game Theory largely fails to incorporate notions of uncertainty critical to the application being considered.
To remedy this, we provide guidance in terms of how to incorporate uncertainty in a model, and what forms of uncertainty are critical to consider in each application area, and how to model the information that is available in each application area.
\end{abstract}



\received{20 February 2007}
\received[revised]{12 March 2009}
\received[accepted]{5 June 2009}

\maketitle

\section{Introduction}
\label{sec:intro}

Effective cybersecurity has become a core pillar of modern life, being necessary for defense, critical infrastructure, and commerce.
This is largely due to the disastrous consequences caused by ineffective cyber defenses.
To give some recent examples in critical infrastructure: the 2021 Colonial Pipeline Breach caused a 6-day shutdown that raised gas prices by approximately 4 cents across 18 states \cite{tsvetanov2021effect}, and 133.8 million patients in the US had protected health information exposed due to hacking or IT-related incidents between 2009 and 2017 \cite{jiang2019evaluation}.
Thus, it is of upmost importance that cybersecurity is taken seriously and made effective.

Although cybersecurity is itself inter-disciplinary and hardening one's cybersecurity posture can take many forms, one important feature to examine is cyber decision-making.
Universally, cyber defenders have finite resources, but enhancing cybersecurity often incurs some cost.
For example, one common cyber defense tool is Intrusion Detection System (IDS), which scans host or network logs for identifying suspicious activities.
Intrinsically, these operations are not free and take up resources (e.g., assigning domain experts to examine IDS alarms) that could otherwise be used for other objectives.
Thus, the fundamental trade-off to cybersecurity decision-making arises: How much should one invest in cybersecurity given the benefit of increased security, but at the cost of system resources?
Answering this trade-off thus becomes a critical question of cybersecurity, and it is clear that ad hoc decision-making will not suffice as any errors may come at a large cost.

To solve this problem, the use of \emph{Game Theory} has been proposed, which studies the interactions of decision-makers.
Game Theory offers two distinct advantages:
first, its solutions are mathematically optimal, so its solutions can be considered disciplined and rigorous.
Secondly, it explicitly considers the strategic dynamics between multiple parties, meaning that the attacker's responses to any given defensive policy are considered.
Both of these attributes are necessary for disciplined cybersecurity decision-making to strike the right balance between security and efficiency.

Despite these advantages over other decision-making regimes, Game Theory has not seen extensive adoption in cybersecurity practice (to our knowledge).
The primary goal of this work is to investigate why this is, characterize exactly what deficits encumber current game-theoretic work, and provide recommendations for future research.

\subsection{Our Contributions}
The main objective of this work is to address two research questions, which we also term our \emph{criteria}:
\begin{itemize}
    \item \emph{Efficacy of Modeling.} To what extent do current game-theoretic models capture the first principles, or the reality, of the underlying cyber interaction? That is, how well do existing models match the real cyber interactions (including both attacker-defender and defender-defender interactions, with the latter demonstrated by data sharing between them)?
    \item \emph{Practicality of Solutions.} Supposing the solutions from game theoretic works are taken as a prescriptive solution for practitioners, can practitioners implement and utilize the solutions? If they cannot, then what elements of the game-theoretic solution prevent it from being utilized? That is, how practical are they to implement?
\end{itemize}
To accomplish this, we make four contributions.
First, we propose a novel framework for systematizing the knowledge of the game-theoretic approach to cybersecurity.
The framework has four pillars: {\em applications} (that have been attempted by the game-theoretic approach), {\em assumptions} (made by the game-theoretic cybersecurity models), {\em models and solution concepts}, and {\em analysis techniques}.


Second, we apply the framework to systematize 80 publications in game-theoretic cybersecurity studies, leading to a clear view of where the state-of-the-art stands with respect to the needs of cybersecurity. In particular, we highlight the importance of considering the {\em game space} level of abstraction, especially as a means of coping with uncertainty. 
However, this perspective is lacking in models in the literature.

Third, we leverage the systematization to evaluate how well existing studies achieve the aforementioned criteria, Efficacy of Modeling, and Practicality of Solutions.
To illuminate the current deficit between game-theoretic cybersecurity and practice, we evaluate our criteria with respect to the framework of \emph{the good}, {\em the bad}, and {\em the ugly}.
We deem the state-of-the-art good if it satisfies our criteria, bad if it satisfies it in a limited way or only minor extensions of existing models are needed, and ugly if significant future work is required to satisfy the criteria.
The ugly classification, in particular, allows us to highlight where the significant unmet needs of cybersecurity are.

Regarding the first criterion, Efficacy of Modeling, we find that game-theoretic works typically leverage a small set of standard game-theoretic models.
Although each model may handle some aspects of the underlying cyber interaction well, all models leave some application-specific needs unmet.
Consistently, one of those needs is to handle uncertainty of the adversary's identity, intentions, and capability, but the majority of traditional game-theoretic approaches do not incorporate this.
This highlights the need for models specifically designed for the needs of cybersecurity.
With respect to the second criterion,  Practicality of Solutions, we find that the traditional notion of a game-theoretic solution is an equilibrium, which encodes a likely emergent behavior of the interaction of independent and selfish decision-makers.
However, calculating an equilibrium requires complete information of the game, including the adversary's capabilities and preferences.
This is obviously problematic in an adversarial cybersecurity setting, creating a significant information deficit between what information is available in a realistic scenario and what is needed to compute a game-theoretic solution.
We 
characterize exactly what pieces of information are required for models to compute their solution and analyze the practicality of those assumptions.

Fourth, we provide recommendations for how future work can overcome the current deficits.
We find that models that are based on real world observations have both better efficacy because the parameters are much more relevant and are more practical to compute because the parameters have a known well-defined source. 
After utilizing this four step process, we summarize our findings and recommendations:
\begin{itemize}
    \item \textbf{The Good.} Game-theoretic approaches universally succeed in providing at least partial guidance in the decision scenario they seek to address.
    \item \textbf{The Bad.} Game-theoretic approaches often fail to provide a complete solution necessary for cybersecurity deployment.
    For example, Intrusion Detection System optimization approaches provide solutions for a static set of parameters but do not formalize parameter recalculation over time.
    This is problematic as this is a necessary consideration for deployment.
    \item \textbf{The Ugly.} One major feature of all cybersecurity problems (especially adversarial scenarios) is the significant uncertainty about the adversary's identity, intent, and capabilities.
    As mentioned above, this is problematic for both criteria.
    This is because if a model that needs precise data about an adversary to function, then it will either output poor answers when poor quality data is input (failing Efficacy of Models) or not be computable at all if no data is input (Failing Practicality of Solutions).
    To remedy this, we recommend future game-theoretic models that with two features: first incorporate uncertainty critical to the application being considered (i.e. the adversary's capability in the context of Advanced Persistent Threats) and leverage the information available in that context (i.e. noisy Intrusion Detection System signals).
    Additionally, we find that building models around cybersecurity metrics and standards such as the Common Vulnerability Scoring System (CVSS) or the Structured Threat Information Expression (STIX) enhances a model's ability to satisfy both criteria.
    This is because they both provide a real world basis for model parameters and are a source of data for said model parameters. 
\end{itemize}

\begin{table*}[!htbp]
    \footnotesize
    \centering
     \begin{tabular}{|l|l|l|l|l|l|l|l|l|l|l|l|l|l|l|l|l|}
    \hline
    \multicolumn{1}{|c}{ } & \multicolumn{4}{|c|}{Applications} & \multirow{1}{*}{\rotH{Assumptions\phantom{llll}}}& \multicolumn{8}{c|}{Game-Theoretic Models} &\multicolumn{2}{c|}{Criteria} 
    \\
    \cline{1-5} \cline{7-16} 
        Paper & APT & IDS & MTD & CTI  & & \rot{Normal} & \rot{Stackelberg} & \rot{Stochastic}  &  \rot{Bayesian} & \rot{Differential} &\rot{Evolutionary} &  \rot{Coalitional}  & \rot{Gestalt} &  \Rot{Efficacy}  \Rot{of Models}& \Rot{Practicality} \Rot{of Models} 
\\ \hline
        \cite{10.1145/2480741.2480742} &  & \cY & ~ &   & & \cY & \cY&\cY   &\cY  & ~ &\cE&\cY& &  ~ & ~ \\ \hline
        \cite{do2017game} & ~ & \cE & ~ & \cY  & \cE & \cY & \cY & \cY  & ~ & \cY & \cY&\cY& &  \cE & \cE \\ \hline
        \cite{kiennert2018survey} & ~ & \cY & ~ & ~ & \cE & \cY & \cY & \cY  & ~ & ~ &&\cE& &\cE   & \cE \\ \hline
        \cite{sengupta2020survey} & \cE & ~ & \cY & ~&  & \cY & \cY & \cY  & \cE & ~&&& ~  & ~ & ~ \\ \hline
        \cite{cho2020toward} & \cE & \cE & \cY & ~  &  & \cY & \cY & \cY  &  \cY &~ &\cE&& &  \cE & \cE \\ \hline
        \cite{pawlick2019game} & \cE & \cY & \cY & ~ & \cE& \cY & \cY & \cE  &  \cY & ~ &&\cE& &  ~ & ~ \\ \hline
        \cite{zhu2021survey} & \cY & ~ & \cY & ~ & \cE& \cY & \cY & \cY &  \cY & & ~ & ~ &&  \cE & \cE \\ \hline
        Ours & \cY & \cY & \cY & \cY  & \cY & \cY & \cY & \cY  & \cY & \cY & \cY & \cY & \cY  & \cY   & \cY \\ \hline
    \end{tabular}
\caption{\label{tab:lit review} Comparison with existing surveys, where a green checkmark (\cY) indicates that a  work thoroughly addresses a topic, and a yellow tilde (\cE) indicates that a work addresses a topic to some extent.}  
\end{table*}

\noindent{\bf Related Work}.
Table~\ref{tab:lit review} compares previous surveys and our systematization of game-theoretic cybersecurity studies. 
Manshaei et al. \cite{10.1145/2480741.2480742} survey five kinds of game-theoretic models for network security including 
IDS optimization.
Do et al.\cite{do2017game} survey six kinds of models, especially Cyber Threat Information (CTI) sharing and IDS optimization, and mention some deficits.
Kiennert et al. \cite{kiennert2018survey} survey three kinds of models for IDS optimization,
while discussing some deficits between these models and cybersecurity practice.
Sengupta et al. \cite{sengupta2020survey} 
survey three kinds of models for Moving Target Defense (MTD).
Cho et al. \cite{cho2020toward} survey four kinds of models for MTD,
while discussing some deficits.
Pawlick et al. \cite{pawlick2019game} 
primarily survey cyber deception models.
Zhu et al. \cite{zhu2021survey} survey deception models, especially in relation to Advanced Persistent Threat (APT) and MTD, while discussing some deficits.
By contrast, our study provides the most rigorous view of the game-theoretic function of existing works.
Second, our criteria provide the most comprehensive view of the current capability of existing models relative to application specific needs, and as well as the deficits preventing adoption of game-theoretic work in the cybersecurity domain. It is worth mentioning that the importance and usefulness of clearly articulating and characterizing the assumptions made by theoretical cybersecurity models, including game-theoretic ones, has been highlighted elsewhere in a broader context \cite{XuSciSec2021SARR}.

\smallskip

\noindent{\bf Paper Outline}. Section \ref{sec:methodology} describes our systematization methodology. Section \ref{sec:applications} discusses four popular cybersecurity applications of game theory.
Section~\ref{sec:assumptions} systematizes assumptions. 
Section~\ref{sec:models} systematizes models and their solutions concepts.
Section \ref{sec:metrics} analyzes existing approaches with respect to our criteria and provides future recommendations. Section \ref{sec:conclusion} concludes the paper. 
Section~\ref{sec:analysis techniques} in the Appendix summarizes popular analysis techniques.


\section{Systematization Methodology and Result Overview}
\label{sec:methodology}

\noindent{\bf Scope}. We focus on game-theoretic models for cybersecurity applications, excluding {\em privacy} applications because they would deserve a systematization on their own. 
We do not investigate the validation of game-theoretic models with real world cybersecurity metrics because the measurement of model parameters is an open problem for future studies.

Our systematization methodology has two parts: {\em systematization framework} and {\em literature selection}.


\subsection{Systematization Framework}
Our systematization framework consists of four pillars:
\begin{itemize}
\item {\bf Applications}: Understanding the cybersecurity applications that have been attempted by the game-theoretic approach is the first step to appreciate its potential. Since there are many applications, we focus on the ones that we deem representative, which suffice for our study but should not be interpreted as underestimating the importance of  other applications.

\item {\bf Assumptions}: It is important to understand the assumptions that have been made by game-theoretic cybersecurity models.
Observing that assumptions in the literature may be implicit,
we articulate the aspects of assumptions that should be explicitly stated in future studies, 
including: (i) those related to {\em model structure}, which specifies the players, their actions, objectives, and action update structure; and (ii) those related to {\em player knowledge}, which specifies what information is known to which players.


\item {\bf Models and Solution Concepts}: 
We systematize the game-theoretic models in the cybersecurity literature into eight types: Normal Form, Stackelberg, Stochastic, Bayesian, Differential,  Evolutionary, Coalitional, and Gestalt games. We also systematize their solution concepts. 

\item {\bf Analysis Techniques}: 
We systematize the techniques that have been used to analyze game-theoretic cybersecurity models. 



\end{itemize} 
These pillars will be elaborated in the subsequent sections, in relation to the literature that will be identified below.
The systematization will yield a clear understanding of the deficits between existing work and the needs of cybersecurity, which leads to exciting future research directions.


\subsection{Literature Selection}\label{sec:lit search}

Relevant studies are scattered in many venues other than traditional 
computer / network / cyber security ones (e.g., Journal of Accounting and Public Policy).
This makes an exhaustive search neither feasible nor warranted. 
Thus, 
we propose conducting search via Google Scholar in two steps: (i) iteratively determining search terms and (ii) using the search terms to conduct the actual literature search.

The first step is conducted iteratively to determine search terms as follows.
We begin by crafting searches for two well-known cybersecurity applications, APT (Advanced Persistent Threat)  and MTD (Moving Target Defense) as well as for survey papers via the search string ``Game Theory in Cybersecurity Survey.''
As the search proceeds, we add new applications (IDS optimization, CTI sharing) as appropriate, and conduct searches based on them by appending the word ``game" to each application (e.g., searching for an application ``X'' via search string ``X game'').
We terminate our application search once we cannot produce a search that finds five new publications in that application area.
The resulting list of search terms is:
``advanced persistent threats game,''
``moving target defense game,''
``honeypot network game,''
``cyber threat information sharing game,''
``intrusion detection system game,'' and
``stealthy game.''

The second step is to use the search terms mentioned above to identify the relevant literature.
For each search term, we conduct two Google Scholar searches, one to emphasize recency and one to highlight influential and highly cited studies.
For each search, we consider the top 20 search results.
In the recency search, we restrict results to the 5-year range 2017-2022 and include all work that meets the following criteria:
published at a reputable 
English venue (which is defined as having a Google Scholar h5-index of at least 25 for technical papers and 
at least 100 for survey papers); addressing a problem in cybersecurity; formulating a game-theoretic model; and presenting solutions with cybersecurity meaning.
For surveys published between 2012-2022 to be considered a game-theorectic survey, we require it to discuss at least two game-theoretic models or solution concepts. 
For the influential work search, we place no restrictions on the publishing date but consider only the top 5 most highly cited studies that meet our criteria. In total, 
our searches find 147 papers, which are manually filtered based on the scoping requirements (in addition to pruning duplicates) using the above criteria, leading to 80 publications for systematization.

\subsection{Systematization Overview}

Table~\ref{tab:abbreviations} describes the abbreviations used in Table~\ref{tab:assumptions}, which
systematizes 80 publications that cover four applications and eight kinds of models.


\begin{table}[!htbp]
    \footnotesize
    \centering
    \begin{tabular}{|l|l|l|}
    \hline
    Term &Full Name   & Reference \\
    \hline
    Convex& Convex Optimization &\cite{boyd2004convex,pedregal2004introduction} \\
    \hline
    DP&Dynamic Programming &Section~\ref{sec:DP}\\
    \hline
    DS & Dominant Strategy & Section~\ref{sec:DS} \\
    \hline
    ESS&Evolutionary Stable Strategy&Section~\ref{sec:evolutionary} \\
    \hline
    Greedy&Greedy Algorithm &\cite{boyd2004convex}\\
    \hline
    HJB &Hamilton-Jacobi-Bellman Equation &\cite{bertsekas2012dynamic,pedregal2004introduction} \\
    \hline
    LCP &Linear Complementarity Programming &Section~\ref{sec:LCP} \\ 
    \hline
    LP&Linear Programming&Section~\ref{sec:LP} \\
    \hline
    NLP&Non-Linear Programming &\cite{pedregal2004introduction} \\
    \hline
    ODE&Ordinary Differential Equation&\cite{pedregal2004introduction} \\
    \hline
    Pontryagin&Pontryagin Minimization Principle &\cite{bertsekas2012dynamic,pedregal2004introduction} \\
    \hline
    Shapely &Shapely Value & \cite{karlin2017game}\\
    \hline
    Variational&Variational Calculus&\cite{pedregal2004introduction} \\
    \hline
    ZD& Zero Determinant Strategy&\cite{adami2013evolutionary} \\
    \hline
    \end{tabular}
    \caption{List of abbreviations used in Table~\ref{tab:assumptions}}
    \label{tab:abbreviations}
\end{table}

\section{Game-Theoretic Cybersecurity Applications}
\label{sec:applications}

Our search indicated that game theory has been applied in four broad categories of cybersecurity problems:
IDS optimization, APT, MTD, and CTI sharing.

\subsection{IDS (Intrusion Detection System) Optimization}
IDS optimization is a general problem because any system that aims to detect attacks can constitute an IDS. IDS optimization often intersects with other applications, such as 
APT and MTD, because they often employ an IDS.
As highlighted in Table \ref{tab:assumptions}, 15 studies fall into this category.

The game-theoretic approach has been applied to address two IDS optimization problems:
{\em false-positives reduction} and {\em resource management}.
(i) False-positive reduction has been modeled as a two-player game between an attacker and an IDS (i.e., defender)
\cite{alpcan2003game,alpcan2006intrusion,alpcan2004game}. Recall that false-positives refer to an IDS wrongly flagging a legitimate activity as an attack.
An IDS makes decisions whether to flag an activity or not, cognizant of the fact that investigating false-positives is resource-consuming. 
(ii) Resource management refers to how an IDS should be allocated or deployed in a network, as demonstrated by two scenarios.
One scenario considers strategic interactions between an attacker and a defender, where the defender chooses systems to equip with an IDS  \cite{huang2020dynamic,niazi2019bayesian,bouhaddi2018efficient,liu2006bayesian,agah2004intrusion,chen2009game,abass2017evolutionary}.
Another scenario is to deploy IDS in a cloud as autonomous players, which may collaborate with each other to increase their effectiveness \cite{abusitta2018trust,li2020glide}.
When IDSes do not naturally collaborate with each other,
incentives may be introduced to encourage cooperation 
(e.g., auction-based incentivization
\cite{guo2018incentive}).

\subsection{APT (Advanced Persistent Threat)}
APT has three characteristics:
(i) the attacker is resourceful;
(ii) the attacker is strategic, with specific targets in mind;
and (iii) the attack is waged in stages where the attacker first gains a foothold in a computer, then expands laterally in the network and escalates their privileges as needed, and finally delivers a payload or exfiltrates data. APT offers rich attacker-defender interactions, leading to 26 studies as highlighted in Table \ref{tab:assumptions}.

There are two main approaches to designing game-theoretic APT models. 
(i) One is initiated by FlipIt \cite{van2013flipit}, where the attacker and the defender ``fight'' for control of a system with limited knowledge about each other's actions.
It can model the stealthiness of APTs, where the defender does not know the state of the computers (owing to the attacker's stealthiness) but must formulate defensive strategies. 
This approach has led to many results \cite{prakash2015empirical,xiao2018attacker,abass2017evolutionary,xiao2017cloud,hu2017defense,zhang2015game,zhang2014stealthy,farhang2016flipleakage,feng2015stealthy,laszka2014flipthem}. 
(ii) Another approach characterizes the multistage nature of APTs via a discrete-time model \cite{basak2019identifying,huang2020dynamic}, continuous-time model \cite{hu2015dynamic,yang2018effective,li2017differential,huang2019adaptive}, or hybrid of both \cite{rass2019cut}.
These studies can model stages of APTs, from reconnaissance to payload delivery.

\subsection{MTD (Moving Target Defense)}
MTD is a class of defensive techniques whereby the defender frequently changes its system configurations to impose uncertainty on the attacker.
MTD has been proposed at multiple levels of abstraction, including network   \cite{sengupta2018moving} 
and host \cite{carter2014game},
but a common trade-off is to balance the cost 
and the security benefit.
The game-theoretic approach often assumes the defender has a finite set of system configurations to switch between, with the objective of finding the optimal way to transition between them to best confuse attackers \cite{tan2019optimal,lei2018incomplete,lei2017optimal,maleki2016markov}.
As such, MTD is often modeled as a Stackelberg game \cite{sengupta2017game,sengupta2018moving,jajodia2018share,feng2017stackelberg}, where the attacker observes the distribution of possible configurations (but not the present configuration) made by 
the defender and acts accordingly.
There are models in a multistage context \cite{carter2014game,chowdhary2018markov,zhu2011robust}, where configurations of previous stages impact future stages.
As highlighted in Table \ref{tab:assumptions}, 28 studies fall into this category.

We treat the deployment of honeypots as one sub-area of MTD. Honeypots refer to fake hosts, resources, or services in the defender's network 
to bait the attacker into interacting with them, causing the attacker to reveal themselves and their exploits.
The research question is centered on
finding an optimal dynamic policy for configuring honeypots to best deceive the attacker.
One approach \cite{basak2019identifying,anwar2022cyber,anwar2020honeypot, horak2019optimizing} is to model the defender's actions are placing honeypots in a network to detect the attacker's movements.
Other approaches consider different complications of honeypots while abstracting the network structure away, such as: the attacker interacts with one \cite{shi2021research} or multiple \cite{diamantoulakis2020game,tian2021honeypot} homogeneous hosts, where the defender is unsure whether an attacker or user is interacting with their system. In these settings, there is a natural need for the defender to leverage the interactions over time to learn the other player(s) characteristics \cite{wang2017strategic,panda2022honeycar,boumkheld2019honeypot,tian2019defense,du2019sdn}.
Additionally, honeypots have been studied in conjunction with IDS \cite{gill2020gtm}. 

\subsection{CTI (Cyber Threat Information) Sharing}\label{sec:CTI}
CTI sharing considers the interactions between entities (e.g., firms) that must decide whether to share their threat intelligence with each other.
Although sharing CTI is an effective way to harden firms' cybersecurity posture, firms frequently opt not to share due to a range of barriers \cite{zibak2019cyber}, such as economic costs and security / privacy concerns.
Due to the strategic nature of firm interactions, the game-theoretic approach has been applied to formulate models to aid the design of systems and incentives to facilitate CTI sharing.
As highlighted in Table \ref{tab:assumptions}, 12 studies fall into this category.

The earliest models are two-firm models \cite{hausken2007information,gao2014game,gao2016differential}, where two firms must decide if to share, when to share, and how much CTI to share.
Then, multiple-firm models \cite{tosh2015evolutionary,tosh2015cyber,tosh2018establishing,vakilinia2017coalitional,vakilinia2019fair} are introduced, while incorporating a trusted third party that handles the aggregation, anonymization, and analysis of CTI data.
Finally, distributed multiple-firm models are studied where firms unilaterally make sharing decisions with other firms in a networked setting~\cite{collins2021paying}. 


\section{Assumptions}
\label{sec:assumptions}

In order to facilitate clear articulation of assumptions made by game-theoretic cybersecurity models, we need to specify the fundamental elements of these models. 
For this purpose, we define two classes of elements: \emph{model structure} and \emph{player knowledge}.
At a high level, model structure defines how a model functions mathematically; player knowledge deals with what a player needs to know in order to use the model to guide decision-making.

To facilitate the systematic articulation of game-theoretic cybersecurity models (including both model structure and player knowledge), we propose considering the three levels of abstraction highlighted in Table~\ref{tab:systemization} and elaborated below:
\begin{itemize}
\item At the {\em current moment} level of abstraction, parameters specify the current information a player is assumed to know, including: the current action $a$,  the current system  state $s$, and the history of actions $h$.
\item At the 
{\em game instance} level of abstraction, parameters 
specify all possible player interactions in a specific game instance, including: the player set $N$, the action set $A$, and the utility function $U$.
\item At the {\em game space} level of abstraction, parameters specify a space of possible game instances, including: the space of player sets $\mathcal{N}$, the space of action sets $\mathcal{A}$, and the space of utility functions $\mathcal{U}$.
\end{itemize}
These three levels of abstraction
form a {\em hierarchy} reflecting the relation $a\in A\in\mathcal{A}$, while noting that $N\in\mathcal{N}$ and $U\in\mathcal{U}$.
\begin{table}[!htbp]
    \small
    \centering
    \begin{tabular}{|l|c|l|}
    \hline
    Level of abstraction & Parameter & Meaning \\
    \hline
    \multirow{3}{*}{Current Moment}&$a$& current action \\
    \cline{2-3}
    ~&$s$& system state \\
    \cline{2-3}
    ~&$h$& history of actions\\
    \hline
    \multirow{3}{*}{Game Instance}&$N$& player set  \\
    \cline{2-3}
    ~&$A$& action set \\
    \cline{2-3}
    ~&$U$& utility function \\
    \hline
    \multirow{3}{*}{Game Space}&$\mathcal{N}$& space of player sets \\
    \cline{2-3}
    ~&$\mathcal{A}$&space of action sets  \\
    \cline{2-3}
    ~&$\mathcal{U}$& space of utility functions \\
    \hline
    \end{tabular}
    \caption{Model parameters in three levels of abstraction \label{tab:abstraction}}
    \label{tab:systemization}
\end{table}

\noindent{\bf Why Should We Consider the Game Space Level of Abstraction?}
Literature models are often specified at the current moment and game instance levels of abstraction, but not the game space level of abstraction.
We advocate for incorporating the game space level of abstraction in future models.
To justify this, let us consider a scenario where the defender is under frequent attacks and is uncertain if the attacks come from a single attacker or multiple attackers; this is common because an attacker can control multiple attacking IP addresses. Existing models cannot cope with this situation of uncertainty, but the hierarchical structure we propose can describe the situation as follows: the defender may know the space of player sets $\mathcal{N}$ (i.e., the attackers that \emph{could be}), but not the realized player set $N$ (i.e., the attackers that \emph{actually are}). This paves a way toward addressing this type of uncertainty.

\subsection{Model Structure}

We systematize model structure via the following elements: {\em player set}, {\em player actions}, {\em player objectives}, and {\em action update structure}.
In Table~\ref{tab:assumptions}, the column entitled
{\em multiple models} indicates whether a work considers multiple variants of a model or not. In the case that it does, we consider what we deem its most impactful result. 

\noindent{\bf Player Set}.
Game-theoretic models are classically defined by a player set $N$ where every player $i\in N$ has decisions to make.
Table~\ref{tab:assumptions} shows that most literature models consider two players (i.e., one attacker and one defender), denoted by $N=\{1,2\}$.
With the newly introduced space of player sets $\mathcal{N}$ for accommodating uncertainty about attackers, we have $\mathcal{N}=\{\{1,2\}\}$ and $N\in \mathcal{N}$ for two-player game instances.



\noindent{\bf Player Actions}.
The action set defines the choices that are available to the players.
Traditionally, player $i$'s actions are given by a set $A_i$ where a single action is denoted by $a_i\in A_i$.
The space of all joint actions is given by $A=A_1\times A_2\times \dots \times A_{|N|}$, and a selection of actions by all players is an \emph{action profile} denoted by $a\in A$.
To model uncertainty associated with an attacker's capabilities, we use the newly introduced space of action sets $\mathcal{A}$
to formalize families of models where attackers have various capabilities.
We define
three types of action sets $A$:
(i) {\em continuous} action sets, where players' actions are subsets of the real numbers; 
(ii) {\em finite}, where players have a finite number of choices;
(iii) {\em mixed}, where finite action sets are often studied in a probabilistic setting such that players select a probability distribution over the action set $A$,
formally defined as $\Delta(A)=\{\sigma\in \mathbb{R}^{|A|}\mid \sum_{a\in A} \sigma(a)=1,\sigma(a)\geq0\textrm{ } \forall a\in A\}$.
Throughout the rest of the paper, we denote a probability distribution over actions by $\sigma$.
In Table~\ref{tab:assumptions} we highlight the action sets of each model as continuous, finite, or mixed $A$.


In cybersecurity, action sets represent players' capabilities, such as scheduling 
scans \cite{yang2018effective,yang2018risk}, cybersecurity investment \cite{gao2014game,gao2016differential}, selecting assets to attack/defend \cite{alpcan2003game}, or taking generic attack/defense actions  \cite{zhu2011robust, huang2020dynamic}.
Consider the situation where an attacker identifies a set of vulnerabilities they can exploit on the defender's system, denoted by 
$A_\textrm{att}=\{\textrm{vulnerability 1},\textrm{vulnerability 2},\ldots\}$, which may not be known to the defender.
To accommodate this uncertainty, we define
$\mathcal{A}_\textrm{att}=\{\textrm{vulnerability set 1},\textrm{vulnerability set 2},\ldots\}$ to describe the situation in which the attacker knows of the existence of different vulnerability sets.


\noindent{\bf Player Objectives}.
The objectives of the players are encoded as \emph{utility functions}.
Each player $i$ has a utility function $U_i:A\rightarrow\mathbb{R}$, which maps every action profile to a real number representing the utility of the action profile to player $i$.
Examples include that a defender suffers a loss in utility when their systems are compromised.
For compactness, we use $U=(U_1,U_2,\dots,U_n)$ to denote the collection of all players' utility functions. We can describe the uncertainty associated with the space of utility functions, $\mathcal{U}$, 
such that each collection of utility functions $U\in \mathcal{U}$ defines all player's objectives in a single game instance.

\begin{landscape}
\scriptsize
    \centering
    \begin{longtable}{|l|l||l|l|l|l|l|l|l|l|l|l||l|l|l|l|l|l||l|l|l|}
\hline
        \multicolumn{2}{|c||}{ } & \multicolumn{10}{c||}{Model Structure 
        } & \multicolumn{6}{c||}{Player Knowledge 
        } &\multicolumn{3}{c|}{Outcome} \\
        \hline
        \rot{Application}&paper & \rot{2-Player} & \rot{Finite $A$} & \rot{Continuous $A$} & \rot{Mixed $A$} & \rot{One-shot $T$} & \rot{Discrete $T$} & \rot{Continuous $T$} & \rot{Sequential $R$} & \rot{Simultaneous $R$} & \rot{Multiple models} & $a$ & $s\cup h$ & $A$ & $N$ & $U_i$ & $U_{-i}$ & Model & Solution  Concept & Analysis Technique \\ \hline
        \multirow{15}{*}{\rot{IDS Optimization}} & \cite{han2019intrusion} & \cY & \cY & \cN & \cY & \cN & \cY & \cN & \cN & \cY & \cN & \cN,\cN & \cN,\cN & \cY,\cY & \cY,\cY & \cY,\cY & \cY,\cY & Normal & Nash & LCP \\ \cline{2-21}
        ~ & \cite{subba2018game} & \cY & \cY & \cN & \cY & \cN & \cN & \cN & \cN & \cY & \cY & \cY,\cY & \cN,\cN & \cY,\cY & \cY,\cY & \cY,\cY & \cY,\cY & Normal & Nash & LCP \\ \cline{2-21}
        ~ & \cite{gothawal2020anomaly} & \cN & \cY & \cN & \cN & \cN & \cY & \cN & \cN & \cY & \cY & \cY,\cY & \cY,\cY & \cY,\cY & \cY,\cY & \cY,\cY & \cY,\cY & Stochastic & Nash & LCP \\ \cline{2-21}
        ~ & \cite{li2017differential} & \cY & \cY & \cN & \cY & \cN & \cN & \cY & \cN & \cY & \cN & \cN,\cN & \cN,\cN & \cY,\cY & \cY,\cY & \cY,\cY & \cY,\cY & Differential & Nash & HJB \\ \cline{2-21}
        ~ & \cite{niazi2019bayesian} & \cY & \cY & \cN & \cY & \cN & \cY & \cN & \cN & \cY & \cN & \cN,\cN & \cY,\cY & \cY,\cY & \cY,\cY & \cY,\cY & \cN,\cY & Bayesian & Bayes-Nash & DS \\ \cline{2-21}
        ~ & \cite{abusitta2018trust} & \cN & \cY & \cN & \cN & \cN & \cY & \cN & \cN & \cY & \cY & \cY,\cY & \cN,\cN & \cY,\cY & \cY,\cY & \cY,\cY & \cN,\cN & Coalition & Coalitions & Numerical \\ \cline{2-21}
        ~ & \cite{li2020glide} & \cN & \cY & \cN & \cY & \cN & \cY & \cN & \cN & \cY & \cN & \cY,\cY & \cN,\cN & \cY,\cY & \cY,\cY & \cY,\cY & \cY,\cY & Normal & Nash & LCP \\ \cline{2-21}
        ~ & \cite{guo2018incentive} & \cN & \cY & \cN & \cY & \cN & \cY & \cN & \cN & \cY & \cY & \cY,\cY & \cN,\cN & \cY,\cY & \cY,\cY & \cY,\cY & \cY,\cY & Evolutionary & ESS & Convex \\ \cline{2-21}
        ~ & \cite{bouhaddi2018efficient} & \cY & \cY & \cN & \cY & \cN & \cY & \cN & \cN & \cY & \cY & \cN,\cN & \cY,\cY & \cY,\cY & \cY,\cY & \cY,\cY & \cN,\cY & Bayesian & Bayes-Nash & DS \\ \cline{2-21}
        ~ & \cite{liu2006bayesian} & \cY & \cY & \cN & \cY & \cN & \cY & \cN & \cN & \cN & \cY & \cN,\cN & \cY,\cY & \cY,\cY & \cY,\cY & \cY,\cY & \cN,\cY & Bayesian & Bayes-Nash & DS \\ \cline{2-21}
        ~ & \cite{agah2004intrusion} & \cY & \cY & \cN & \cY & \cN & \cY & \cN & \cN & \cY & \cY & \cN,\cN & \cN,\cN & \cY,\cY & \cY,\cY & \cY,\cY & \cY,\cY & Normal & Nash  & Q Learning \\ \cline{2-21}
        ~ & \cite{alpcan2004game} & \cY & \cY & \cN & \cY & \cN & \cY & \cN & \cN & \cY & \cY & \cE,\cN & \cN,\cN & \cY,\cY & \cY,\cY & \cY,\cY & \cY,\cY & Extensive & Nash & Convex \\ \cline{2-21}
        ~ & \cite{alpcan2003game} & \cY & \cY & \cN & \cY & \cY & \cY & \cN & \cY & \cN & \cY & \cE,\cN & \cN,\cN & \cY,\cY & \cY,\cY & \cY,\cY & \cY,\cY & Stackelberg & Nash & Shapley \\ \cline{2-21}
        ~ & \cite{alpcan2006intrusion} & \cY & \cY & \cN & \cY & \cN & \cY & \cN & \cY & \cN & \cY & \cE,\cN & \cN,\cN & \cY,\cY & \cY,\cY & \cY,\cY & \cY,\cY & Normal & Nash & Q learning \\ \cline{2-21}
        ~ & \cite{chen2009game} & \cN & \cY & \cN & \cY & \cY & \cY & \cN & \cN & \cY & \cY & \cN,\cN & \cN,\cN & \cY,\cY & \cY,\cY & \cY,\cY & \cY,\cY & Normal & Nash & LCP \\ \hline
        \multirow{26}{*}{\rot{APT}} & \cite{rass2017defending} & \cY & \cY & \cN & \cY & \cY & \cY & \cN & \cN & \cY & \cY & \cN,\cN & \cN,\cN & \cY,\cY & \cY,\cY & \cE,\cE & \cE,\cE & Normal & Nash & Prob. Theory \\ \cline{2-21}
        ~ & \cite{yang2018effective} & \cY & \cN & \cY & \cN & \cN & \cN & \cY & \cN & \cY & \cN & \cN,\cN & \cN,\cN & \cY,\cY & \cY,\cY & \cY,\cY & \cY,\cY & Differential & Nash & Pontryagin \\ \cline{2-21}
        ~ & \cite{zhu2018multi} & \cY & \cY & \cN & \cY & \cN & \cY & \cN & \cN & \cY & \cN & \cN,\cN & \cY,\cY & \cY,\cY & \cY,\cY & \cY,\cY & \cY,\cY & Gestalt & Gestalt & DP \\ \cline{2-21}
        ~ & \cite{xiao2018attacker} & \cY & \cN & \cY & \cN & \cN & \cN & \cY & \cN & \cY & \cN & \cN,\cN & \cN,\cN & \cY,\cY & \cY,\cY & \cY,\cY & \cY,\cY & Flipit & Nash & Prospect \\ \cline{2-21}
        ~ & \cite{abass2017evolutionary} & \cY & \cN & \cY & \cN & \cN & \cN & \cY & \cN & \cY & \cN & \cN,\cN & \cN,\cN & \cY,\cY & \cY,\cY & \cY,\cY & \cY,\cY & Flipit & ESS & Pontryagin \\ \cline{2-21}
        ~ & \cite{xiao2017cloud} & \cY & \cN & \cY & \cN & \cN & \cN & \cY & \cN & \cY & \cY & \cN,\cN & \cN,\cN & \cY,\cY & \cY,\cY & \cY,\cY & \cY,\cY & Flipit & Nash & Convex \\ \cline{2-21}
        ~ & \cite{min2018defense} & \cY & \cN & \cY & \cN & \cY & \cY & \cN & \cY & \cN & \cN & \cN,\cY & \cN,\cN & \cY,\cY & \cY,\cY & \cY,\cY & \cY,\cY & Normal & Nash & Q learning \\ \cline{2-21}
        ~ & \cite{yang2018risk} & \cY & \cN & \cY & \cN & \cN & \cN & \cY & \cN & \cY & \cN & \cN,\cN & \cY,\cY & \cY,\cY & \cY,\cY & \cY,\cY & \cY,\cY & Differential & Nash  & Greedy \\ \cline{2-21}
        ~ & \cite{huang2019adaptive} & \cY & \cY & \cN & \cY & \cN & \cY & \cN & \cN & \cY & \cN & \cN,\cN & \cY,\cY & \cY,\cY & \cY,\cY & \cY,\cY & \cN,\cY & Bayesian & Bayes-Nash & DP \\ \cline{2-21}
        ~ & \cite{huang2018analysis} & \cY & \cY & \cN & \cY & \cN & \cY & \cN & \cN & \cY & \cY & \cN,\cN & \cY,\cY & \cY,\cY & \cY,\cY & \cY,\cY & \cN,\cN & Bayesian & Bayes-Nash & DP \\ \cline{2-21}
        ~ & \cite{hu2017defense} & \cY & \cY & \cN & \cY & \cN & \cN & \cY & \cN & \cY & \cY & \cE,\cN & \cN,\cN & \cY,\cY & \cY,\cY & \cY,\cY & \cY,\cY & Flipit & Nash & Convex \\ \cline{2-21}
        ~ & \cite{huang2020dynamic} & \cY & \cY & \cN & \cY & \cN & \cY & \cN & \cN & \cY & \cN & \cN,\cN & \cY,\cY & \cY,\cY & \cY,\cY & \cY,\cY & \cY,\cY & Bayesian & Bayes-Nash & DP \\ \cline{2-21}
        ~ & \cite{hu2015dynamic} & \cN & \cN & \cY & \cN & \cY & \cN & \cY & \cN & \cY & \cN & \cN,\cN & \cY,\cY & \cY,\cY & \cY,\cY & \cY,\cY & \cY,\cY & Differential & Nash & Pontryagin \\ \cline{2-21}
        ~ & \cite{pawlick2015flip} & \cN & \cY & \cY & \cY & \cY & \cY & \cY & \cY & \cY & \cN & \cY,\cY & \cN,\cN & \cY,\cY & \cY,\cY & \cY,\cY & \cY,\cY & Gestalt & Gestalt & Inequalities \\ \cline{2-21}
        ~ & \cite{basak2019identifying} & \cN & \cY & \cN & \cN & \cN & \cY & \cN & \cN & \cY & \cN & \cN,\cN & \cY,\cY & \cY,\cY & \cY,\cY & \cY,\cY & \cN,\cY & Bayesian & Strategies & Numerical \\ \cline{2-21}
        ~ & \cite{aydeger2021strategic} & \cY & \cY & \cN & \cN & \cN & \cY & \cN & \cN & \cY & \cN & \cY,\cN & \cY,\cY & \cY,\cY & \cY,\cY & \cY,\cY & \cN,\cY & Bayesian & Bayes-Nash & DS \\ \cline{2-21}
        ~ & \cite{rass2019cut} & \cY & \cY & \cN & \cY & \cN & \cY & \cN & \cN & \cN & \cN & \cN,\cY & \cY,\cY & \cY,\cY & \cY,\cY & \cY,\cY & \cN,\cY & Bayesian & Bayes-Nash & Numerical \\ \cline{2-21}
        ~ & \cite{umsonst2018game} & \cY & \cN & \cY & \cN & \cY & \cY & \cN & \cY & \cN & \cN & \cN,\cY & \cN,\cN & \cY,\cY & \cY,\cY & \cY,\cY & \cY,\cY & Stackelberg & Stackelberg & Convex \\ \cline{2-21}
        ~ & \cite{fotiadis2020constrained} & \cY & \cN & \cY & \cN & \cN & \cN & \cY & \cN & \cY & \cY & \cN,\cN & \cY,\cY & \cY,\cY & \cY,\cY & \cY,\cY & \cY,\cY & Differential & Nash & Pontryagin \\ \cline{2-21}
        ~ & \cite{anwar2017dynamic} & \cY & \cY & \cN & \cY & \cN & \cY & \cN & \cN & \cY & \cN & \cN,\cN & \cY,\cY & \cY,\cY & \cY,\cY & \cY,\cY & \cY,\cY & Stochastic & Nash & Value Iteration \\ \cline{2-21}
        ~ & \cite{zhang2015game} & \cY & \cN & \cY & \cN & \cN & \cN & \cY & \cN & \cY & \cY & \cN,\cY & \cN,\cN & \cY,\cY & \cY,\cY & \cY,\cY & \cY,\cY & Flipit & Stackelberg & LP \\ \cline{2-21}
        ~ & \cite{zhang2014stealthy} & \cY & \cN & \cY & \cN & \cN & \cN & \cY & \cN & \cY & \cN & \cN,\cY & \cN,\cN & \cY,\cY & \cY,\cY & \cY,\cY & \cY,\cY & Flipit & Nash & Convex \\ \cline{2-21}
        ~ & \cite{farhang2016flipleakage} & \cY & \cN & \cY & \cN & \cN & \cN & \cY & \cN & \cY & \cN & \cN,\cY & \cN,\cN & \cY,\cY & \cY,\cY & \cY,\cY & \cY,\cY & Flipit & Nash & Inequalities \\ \cline{2-21}
        ~ & \cite{van2013flipit} & \cY & \cN & \cY & \cN & \cN & \cN & \cY & \cN & \cY & \cY & \cN,\cN & \cN,\cN & \cY,\cY & \cY,\cY & \cY,\cY & \cY,\cY & Flipit & Nash & DS \\ \cline{2-21}
        ~ & \cite{feng2015stealthy} & \cN & \cN & \cY & \cN & \cN & \cN & \cY & \cN & \cY & \cN & \cN,\cN & \cN,\cN & \cY,\cY & \cY,\cY & \cY,\cY & \cY,\cY & Flipit & Subgame & Calculus \\ \cline{2-21}
        ~ & \cite{laszka2014flipthem} & \cY & \cN & \cY & \cN & \cN & \cY & \cN & \cN & \cY & \cN & \cN,\cY & \cN,\cN & \cY,\cY & \cY,\cY & \cY,\cY & \cY,\cY & Flipit & Nash & LP \\ \hline
        ~ & \cite{tan2019optimal} & \cY & \cY & \cN & \cY & \cN & \cY & \cN & \cN & \cY & \cN & \cN,\cN & \cY,\cY & \cY,\cY & \cY,\cY & \cY,\cY & \cY,\cY & Stochastic & Nash & NLP \\ \cline{2-21}
        ~ & \cite{feng2017signaling} & \cY & \cY & \cN & \cY & \cY & \cY & \cN & \cY & \cN & \cN & \cN,\cY & \cN,\cN & \cY,\cY & \cY,\cY & \cY,\cY & \cY,\cY & Signal & Stackelberg & Convex \\ \cline{2-21}
        ~ & \cite{lei2018incomplete} & \cY & \cY & \cN & \cY & \cN & \cY & \cN & \cY & \cN & \cN & \cY,\cN & \cY,\cY & \cY,\cY & \cY,\cY & \cY,\cY & \cY,\cY & Stochastic & Nash & Reduction \\ \cline{2-21}
        ~ & \cite{sengupta2017game} & \cY & \cY & \cN & \cY & \cY & \cY & \cN & \cY & \cN & \cN & \cN,\cY & \cN,\cN & \cY,\cY & \cY,\cY & \cY,\cY & \cY,\cY & Stackelberg & Stackelberg  & Convex \\ \cline{2-21}
        ~ & \cite{feng2017stackelberg} & \cY & \cY & \cN & \cY & \cN & \cY & \cN & \cY & \cN & \cN & \cN,\cY & \cN,\cN & \cY,\cY & \cY,\cY & \cY,\cY & \cY,\cY & Stackelberg & Stackelberg & Induction \\ \cline{2-21}
        \multirow{20}{*}{\rot{MTD}} & \cite{wang2019moving} & \cY & \cY & \cN & \cY & \cY & \cY & \cN & \cN & \cY & \cN & \cN,\cN & \cY,\cY & \cY,\cY & \cY,\cY & \cY,\cY & \cY,\cY & Normal & Payoff & ZD \\ \cline{2-21}
        ~ & \cite{maleki2016markov} & \cY & \cY & \cN & \cY & \cN & \cN & \cY & \cY & \cN & \cY & \cN,\cN & \cY,\cY & \cY,\cY & \cY,\cY & \cY,\cY & \cN,\cN & Stochastic & statistics & Markov \\ \cline{2-21}
        ~ & \cite{sengupta2018moving} & \cY & \cY & \cN & \cY & \cY & \cY & \cN & \cY & \cN & \cN & \cN,\cY & \cN,\cN & \cY,\cY & \cY,\cY & \cY,\cY & \cY,\cY & Stackelberg & Stackelberg  & LP \\ \cline{2-21}
        ~ & \cite{lei2017optimal} & \cY & \cY & \cN & \cY & \cY & \cY & \cN & \cN & \cY & \cN & \cN,\cN & \cY,\cY & \cY,\cY & \cY,\cY & \cY,\cY & \cY,\cY & Stochastic & Nash  & NLP \\ \cline{2-21}
        ~ & \cite{chowdhary2018markov} & \cY & \cY & \cN & \cY & \cN & \cY & \cN & \cY & \cN & \cN & \cN,\cY & \cY,\cY & \cY,\cY & \cY,\cY & \cY,\cY & \cY,\cY & Stochastic & Stackelberg  & Q learning \\ \cline{2-21}
        ~ & \cite{prakash2015empirical} & \cY & \cY & \cN & \cN & \cN & \cN & \cY & \cN & \cY & \cN & \cY,\cY & \cY,\cY & \cY,\cY & \cY,\cY & \cY,\cY & \cY,\cY & Flipit & Nash & Numerical \\ \cline{2-21}
        ~ & \cite{zhu2013game} & \cY & \cY & \cN & \cY & \cN & \cY & \cN & \cN & \cY & \cN & \cN,\cN & \cY,\cY & \cN,\cN & \cN,\cN & \cY,\cY & \cY,\cY & Differential & Nash & Convex \\ \cline{2-21}
        ~ & \cite{wang2017strategic} & \cY & \cY & \cN & \cY & \cN & \cY & \cN & \cN & \cY & \cN & \cN,\cN & \cN,\cN & \cY,\cY & \cY,\cY & \cY,\cY & \cY,\cY & Extensive & Bayes-Nash & DS \\ \cline{2-21}
        ~ & \cite{shi2021research} & \cN & \cY & \cN & \cY & \cN & \cN & \cY & \cY & \cN & \cN & \cY,\cY & \cN,\cN & \cY,\cY & \cY,\cY & \cY,\cY & \cY,\cY & Evolutionary & ESS & Lyapunov \\ \cline{2-21}
        ~ & \cite{anwar2020honeypot} & \cY & \cY & \cN & \cY & \cN & \cY & \cN & \cN & \cY & \cN & \cN,\cN & \cN,\cN & \cY,\cY & \cY,\cY & \cY,\cY & \cY,\cY & Normal & Nash & LP \\ \cline{2-21}
        ~ & \cite{horak2019optimizing} & \cY & \cY & \cN & \cY & \cN & \cY & \cN & \cN & \cY & \cN & \cN,\cN & \cY,\cY & \cY,\cY & \cY,\cY & \cY,\cY & \cY,\cY & Stochastic & Nash & Convex \\ \cline{2-21}
        ~ & \cite{tian2019defense} & \cY & \cY & \cN & \cY & \cN & \cY & \cN & \cN & \cY & \cN & \cN,\cN & \cY,\cY & \cY,\cY & \cY,\cY & \cY,\cY & \cY,\cY & Extensive & Bayes-Nash & DS \\ \cline{2-21}
        ~ & \cite{diamantoulakis2020game} & \cY & \cY & \cN & \cY & \cN & \cY & \cN & \cN & \cY & \cY & \cY,\cY & \cY,\cY & \cY,\cY & \cY,\cY & \cY,\cY & \cY,\cY & Normal & Bayes-Nash & Convex \\ \cline{2-21}
        ~ & \cite{tian2021honeypot} & \cY & \cN & \cY & \cY & \cN & \cN & \cY & \cN & \cY & \cN & \cN,\cN & \cN,\cN & \cY,\cY & \cY,\cY & \cY,\cY & \cY,\cY & Evolutionary & ESS & ODE \\ \cline{2-21}
        ~ & \cite{gill2020gtm} & \cY & \cY & \cN & \cY & \cN & \cY & \cN & \cN & \cY & \cN & \cN,\cN & \cN,\cN & \cY,\cY & \cY,\cY & \cY,\cY & \cY,\cY & Normal & Nash & LCP \\ \cline{2-21}
        ~ & \cite{boumkheld2019honeypot} & \cY & \cY & \cN & \cY & \cY & \cY & \cN & \cY & \cN & \cN & \cN,\cN & \cY,\cY & \cY,\cY & \cY,\cY & \cY,\cY & \cY,\cY & Bayesian & Bayes-Nash & LCP \\ \cline{2-21}
        ~ & \cite{panda2022honeycar} & \cY & \cY & \cN & \cY & \cN & \cY & \cN & \cN & \cY & \cY & \cN,\cN & \cN,\cN & \cY,\cY & \cY,\cY & \cY,\cY & \cY,\cY & Normal & Nash & NLP \\ \cline{2-21}
        ~ & \cite{anwar2022cyber} & \cY & \cY & \cN & \cY & \cY & \cY & \cN & \cN & \cY & \cN & \cN,\cN & \cN,\cN & \cY,\cY & \cY,\cY & \cY,\cY & \cY,\cY & Normal & Nash & Convex  \\ \cline{2-21}
        ~ & \cite{wang2022learning} & \cN & \cY & \cN & \cY & \cN & \cN & \cY & \cY & \cN & \cY & \cN & \cN & \cY & \cY & \cY & \cN & Normal & Contract & Induction \\ \cline{2-21}
        ~ & \cite{du2019sdn} & \cY & \cY & \cN & \cN & \cY & \cY & \cN & \cY & \cN & \cN & \cN,\cE & \cN,\cN & \cY,\cY & \cY,\cY & \cY,\cY & \cY,\cY & Bayesian & Bayes-Nash & LCP \\ \cline{2-21}
        ~ & \cite{anwar2022honeypot} & \cY & \cY & \cN & \cY & \cN & \cY & \cN & \cN & \cY & \cY & \cN,\cN & \cN,\cN & \cY,\cY & \cY,\cY & \cY,\cY & \cY,\cY & Normal & Nash & LP \\ \cline{2-21}
        ~ & \cite{pibil2012game} & \cY & \cY & \cN & \cY & \cY & \cY & \cN & \cY & \cN & \cN & \cN,\cN & \cN,\cN & \cY,\cY & \cY,\cY & \cY,\cY & \cY,\cY & Extensive & Stackelberg & LP \\ \cline{2-21}
        ~ & \cite{la2016deceptive} & \cY & \cY & \cN & \cY & \cN & \cY & \cN & \cY & \cN & \cN & \cN,\cN & \cY,\cY & \cY,\cY & \cY,\cY & \cY,\cY & \cY,\cY & Bayesian & Bayes-Nash  & DS \\ \hline
        \multirow{12}{*}{\rot{CTI Sharing}} & \cite{vakilinia2017coalitional} & \cN & \cY & \cN & \cN & \cN & \cY & \cN & \cN & \cY & \cN & \cN & \cN & \cY & \cY & \cY & \cY & Coalition & Stability & Shapley \\ \cline{2-21}
        ~ & \cite{vakilinia2018coalitional} & \cN & \cY & \cN & \cN & \cN & \cY & \cN & \cN & \cY & \cN & \cN & \cN & \cY & \cY & \cY & \cY & Coalition & Insurance & Convex \\ \cline{2-21}
        ~ & \cite{nagurney2017multifirm} & \cN & \cN & \cY & \cN & \cY & \cY & \cN & \cN & \cY & \cY & \cN & \cN & \cY & \cY & \cY & \cY & Differential & Nash  & Variational \\ \cline{2-21}
        ~ & \cite{tosh2015cyber} & \cN & \cN & \cY & \cN & \cY & \cY & \cN & \cN & \cY & \cN & \cN & \cN & \cY & \cY & \cY & \cY & Differential & Nash  & Convex \\ \cline{2-21}
        ~ & \cite{kamhoua2015cyber} & \cN & \cY & \cN & \cN & \cY & \cY & \cN & \cY & \cY & \cN & \cN & \cN & \cY & \cY & \cY & \cY & Normal & Nash & DS \\ \cline{2-21}
        ~ & \cite{hausken2007information} & \cY & \cN & \cY & \cN & \cN & \cN & \cY & \cN & \cY & \cN & \cY & \cN & \cY & \cY & \cY & \cY & Differential & Nash & ODE \\ \cline{2-21}
        ~ & \cite{gao2014game} & \cY & \cN & \cY & \cN & \cN & \cN & \cY & \cN & \cY & \cN & \cY & \cN & \cY & \cY & \cY & \cY & Differential & Nash & ODE \\ \cline{2-21}
        ~ & \cite{gao2016differential} & \cY & \cN & \cY & \cN & \cN & \cN & \cY & \cN & \cY & \cN & \cY & \cN & \cY & \cY & \cY & \cY & Differential & Nash & ODE \\ \cline{2-21}
        ~ & \cite{tosh2015evolutionary} & \cN & \cY & \cY & \cY & \cN & \cN & \cY & \cN & \cY & \cN & \cY & \cN & \cY & \cY & \cY & \cY & Evolutionary & ESS & ODE \\ \cline{2-21}
        ~ & \cite{tosh2018establishing} & \cN & \cY & \cY & \cY & \cN & \cN & \cY & \cN & \cY & \cN & \cY & \cN & \cY & \cY & \cY & \cY & Evolutionary & ESS & ODE \\ \cline{2-21}
        ~ & \cite{collins2021paying} & \cN & \cY & \cN & \cN & \cN & \cY & \cN & \cN & \cY & \cN & \cY & \cN & \cY & \cY & \cY & \cY & Normal & Nash & LCP \\ \hline
        \caption{ Systematization of the game-theoretic approach to cybersecurity in three perspectives: player knowledge, model structure, and outcomes, where
        each row refers to a paper and each column describes an attribute of a paper.
        In the player knowledge columns, a green checkmark (\cY) means that a player needs to know the corresponding model parameter; for a cell containing two symbols, the first (second) symbol indicates if the defender (attacker) needs to know the parameter in question.
        In the model structure columns, a green checkmark (\cY) means whether the listed attribute is described in the paper in question.
        The outcomes columns highlights the model type, the analysis technique, and the solution concept that are used in the paper in question, where the abbreviations are defined in Table~\ref{tab:abbreviations}. 
    \label{tab:assumptions}}
    \end{longtable}
        
\end{landscape}

\noindent{\bf Action Update Structure}.
For fixed players, actions and utility functions, the outcome of a model can vary dramatically depending on the {\em timescale} the players play on and the {\em order of play}, which we collectively call {\em action update structure}. 
We formalize the timescale of a game as set $T\subseteq \mathbb{R}$ such that any $t\in T$ is an opportunity for at least one player 
to make an action in order.
The order of play describes the order in which players make their actions (e.g., alternating or making actions simultaneously).
Formally, we describe the order of play via a {\em revision} rule $R:T\rightarrow 2^N$, which takes the time of a game as input and returns a subset of players that may update their action, where $2^N$ denotes the power set of $N$.
For example, if at some revision time $t\in T$ we have $R(t)=\{1\}$, then player 1 may update their action at a time $t$.
Although many complex definitions of $R$ are possible, two common ones are:
\begin{itemize}
\item \emph{Sequential play}: Players alternate updating their action, meaning $R(t)=\{i\}$ for some $i\in N$ and $R(t)\neq R(t+1)$ in the discrete time case.
\item \emph{Simultaneous play}: Players update their actions simultaneously, meaning $R(t)=N$ for all $t\in T$.
\end{itemize}
Using the preceding formulation, we categorize a model as 
having one-shot, discrete, or continuous $T$ and 
Sequential or Simultaneous $R$, as categorized in Table~\ref{tab:assumptions} and elaborated below.

First, for one-shot $T$ or \emph{one-shot game}, which is a special case of the discrete-time model, each player has one revision opportunity.
In the sequential move setting,
we have $T=\{1,2,3,\dots,|N|\}$ and $R(t)=\{i\}$ for $t\in T$ and $i\in N$ and $R(t)\neq R(t+1)$;
in the simultaneous play setting, we have $T=\{1\}$ accompanied by revision rule $R(1)=N$.
Second, a discrete timescale
is often used when players have sufficient time to make decisions, such as MTD \cite{sengupta2017game,sengupta2018moving} where the defender first calculates their randomization scheme and the attacker has time to observe the resulting distribution before taking action.
Third, on a continuous timescale, for each revision opportunity $t\in T$ there exist nearby revision opportunities given either by $t+\epsilon$ or $t-\epsilon$ for any arbitrarily small $\epsilon>0$.
For example, the FlipIt \cite{van2013flipit,pawlick2015flip,laszka2014flipthem} family of models use a continuous timescale, where the attacker and defender can take action at any time.
\subsection{Player Knowledge}
\label{sec:player knowledge}


To help articulate player knowledge while accommodating conventions in the literature, we specify player knowledge using the following two notions (which cut across the four aspects of the model structure).
\begin{itemize}
\item \emph{History of play} is the sequence of action profiles up to the current time $t$, denoted by $h=(a^1,a^2,\dots,a^{t-1})$ in the discrete-time model, where $a^\ell\in A$ is the action profile at time $\ell$.
Players learn the history of play 
as it progresses and can leverage the history to infer how other players will behave in the future.
\item {\em State space}, $S$, encodes the current environment where the players interact with each other.
At any time $t$, the game is in a state $s\in S$ and transitions between states at each time step as a function of the player's actions.
Examples of states in cybersecurity include the security state of a network (e.g., which computers are compromised and which are not).
\end{itemize}
Formally, 
we denote the complete {\em player knowledge} by $K^c=\{\mathcal{N},\mathcal{A},\mathcal{U},N,A,U,a,T,R,S,s,h\}$, where notations are defined above.
We write $K_i\subseteq K^c$ to denote the knowledge of player $i$. 
Our examination shows that {\em no single} game-theoretic cybersecurity model in the literature has explicitly stated all these aspects of $K_i$; this severely limits the usefulness of these models because any practical application would require that $K_i$ is measured.

We highlight the six most salient elements of player knowledge 
in Table~\ref{tab:assumptions},
where $s\cup h$ indicates that the players have knowledge of the state or history (while recalling that both $s$ and $h$ are functions of previous plays).
In each of the six player-knowledge columns, there is a pair of symbols, where the first symbol pertains to the defender and the second to the attacker.
We note that
most references have two \cY's or \cN's, where \cY\ means the player has knowledge of the model parameter 
and
\cN\ means the player does not.
For example, in the $U_{-i}$ column, (\cN,\cY) means that the defender does not know the attacker's utility function, but the attacker knows the defender's utility function.
From the defender's point of view, the interpretation of a \cY\ is:
\begin{itemize}
    \item $a$: defender observes the attacker's action,
    \item $s\cup h$: defender observes the current state or history of actions,
    \item $N$: defender knows 
    all players, 
    \item $A$: defender can measure all capabilities of all attackers, 
    \item $U_i$: defender knows their own utility function,
    \item $U_{-i}$: defender knows the other player's 
    utility function.
\end{itemize}
Note that Table~\ref{tab:assumptions} only highlights the six kinds of player knowledge 
because they have the highest cybersecurity significance and we have limited space in terms of the table size allowed.
This does not undermine the integrity of the present study for two reasons:
(i) among the other model parameters that are not selected, $\mathcal{N},\mathcal{A},\mathcal{U},R,S$ 
can be skipped because they are assumed by {\em all} the cited references 
as known to the players; (ii)
the timescale $T$ is omitted because its cybersecurity significance is not discussed in the literature (e.g., the plausibility of using a continuous- or discrete-time model).

\subsection{Common Literature Assumptions}

As highlighted in Table \ref{tab:assumptions}, the following assumptions related to {\em model structure} are often made in the literature.
(i) The vast majority of models consider only two-player games. This is problematic; attackers in the real world can interact with each other, either directly or indirectly, which complicates the situation for the defender.
(ii) Models usually consider a finite action set with mixed strategies.
Although mathematically elegant, the cybersecurity implications of probabilistic actions are not always well understood or justified.
(iii) Models often use discrete-time interactions; this can be problematic when attackers and defenders do not operate on the same timescale (i.e., automated attack vs. manual defense).

The following assumptions related to {\em player knowledge} are often made in the literature.
(i) Players can observe the current state $s$ as shown in the $s\cup h$ column. This strong assumption is often difficult to satisfy because it is often challenging to determine in practice, for example, whether a computer is compromised or not. 
(ii) Players often know each other's action set as shown in the $A$ column. This is difficult to justify in the real world. 
(iii) All players know the number and identities of all other players, as shown in the $N$ column.
This strong assumption demonstrates their limited capability in describing uncertainty from this aspect.
(iv) Players often are assumed to know each other's utility function as shown by the many \cY's in the $U_i$ and $U_{-i}$ columns of Table \ref{tab:assumptions}.
In the real world, it would be rare that a defender knows the attacker's objectives or utility functions.




\section{Models and Solution Concepts}
\label{sec:models}
\label{sec:solutions}



\noindent{\bf Classification of Games}.
We divide existing game-theoretic cybersecurity models into eight categories based on how a model is structured, 
and what player knowledge is assumed.
(i) In {\em Normal Form games}, players have no knowledge of the other players' behavior and must base their actions solely off their own utility functions.
(ii) In {\em Stackelberg games}, players make actions according to a given order of play, where one player moves first and the other player observes and reacts to the first player's action. 
{\em Extensive form games} are a generalization  of  Stackelberg games, but we highlight Stackelberg games because of their prominence in the literature.
(iii) In {\em Stochastic games}, there is an environment that changes based on the behavior of players, impacts their utility, and is observable. We treat {\em FlipIt} as a special case of stochastic games.
(iv) In {\em Bayesian games}, players maintain beliefs about objectives of other players, and update their beliefs based on the history of actions they have observed. {\em  Signaling games} are a special case of Bayesian games.
(v) In {\em Differential games}, both time and state can be continuous rather than discrete.
(vi) In {\em Evolutionary games}, actions are interpreted as populations of different species and require a more robust equilibrium than the classical Nash equilibrium.
(vii) In {\em Coalitional games},  players decide on how to form coalitions or cooperative groups among themselves.
(viii) In {\em Gestalt games}, different models are used at different stages, where outcomes of previous stages impact following stages.
We highlight the game-theoretic model of each reference in Table~\ref{tab:assumptions} under the Model column, while using the refined categorization when applicable (e.g., FlipIt rather than Stochastic game is listed because FlipIt is a special case of Stochastic game).



\noindent{\bf Solution Concepts: Predictive vs. Prescriptive Equilibrium and Implications}. A game-theoretic study typically derives an equilibrium as its solution concept, which describes the players' behavior in the model.
In classical game theory, solution concepts are typically treated as \emph{predictive}. For example, a Nash equilibrium is used to characterize the emergent behavior resulting from players' uncoordinated but strategic plays.
In this view, individual players are never required to compute an equilibrium, which
is instead treated as an emergent outcome of strategic interactions that can be ``learned'' collectively by the players~\cite{Pradelski2012,Papadimitriou2016}.
That is, an equilibrium is derived only to understand the behavior at which the players will settle; during play, players only need to evaluate their utility function and update their action accordingly. 

In contrast, game-theoretic cybersecurity studies need to
treat solution concepts as \emph{prescriptive} because a defender is expected to compute (for example) a Nash equilibrium offline and then deploy the associated strategy in practice. 
This increases the knowledge required by the defender, as they are no longer learning their action by interaction but instead deriving it based on a model. For example, having knowledge of the player set $N$ is non-trivial in practice but is needed to compute equilibrium behavior. Since most types of equilibrium cannot be computed without the entire game specification~\cite{Hart2010}, this shift from the predictive paradigm to the prescriptive paradigm may create a {\em large gap}
between the knowledge that the players need to have in order to compute their strategies and the knowledge that is available in a real world setting.
The implication of this predictive vs. prescriptive view is why we specify {\em player knowledge} as mentioned above, namely that player knowledge should include all information that is needed to compute the equilibrium.

\subsection{Models and Their Solution Concepts}

\subsubsection{Normal Form Games} \label{ssec:normal form game}
These are simultaneous move, one-shot, finite-action, $|N|$-player games with a utility function $u_i:A\rightarrow \mathbb{R}$ that can be represented by a $|A_1|\times|A_2|\times \dots \times |A_{|N|}|$ matrix.
It can be used to study, for example, IDS optimization \cite{alpcan2003game,alpcan2006intrusion,agah2004intrusion}, where the IDS (defender)
is to flag or not flag the attacker's behavior while considering the cost incurred by false-positives.
To see an example of this, we give an example proposed in the seminal work on IDS optimization \cite{alpcan2003game}:
\begin{center}
\begin{tabular}{ |r|rr| } 
 \hline
 & Attack & Wait \\ 
 \hline
 Alarm & $-1,-1$ & $-1,0$ \\ 
 Wait & $-5,5$ & $0,0$ \\ 
 \hline
\end{tabular}
\end{center}
where the rows represent the defender's action (either set off an IDS alarm or do nothing) and the columns are the attacker's action (attack or do nothing).
The first number in each quadrant refers to the row player's (defender) utility and the second number refers to the column player's (attacker) utility.
It can be seen that if the attacker waits, the defender prefers to also wait or they will incur a small cost for setting off a false alarm.
However, if the attacker strikes, the defender greatly prefers to raise the alarm.
Conversely, the attacker wishes to strike when the alarm is not raised and otherwise prefers to wait as a small cost is incurred when they attack and the defender correctly raises the alarm.

These games do not intrinsically prescribe any game structure in terms of how players make decisions.
Instead, their \emph{equilibria} are 
defined as action profiles satisfying various joint conditions on player utilities.
The most studied solution concept is {\em Nash equilibrium}, which captures the self-interested behavior that no individual player can change their action to increase their utility.
Formally, $a^*\in A$ is a Nash equilibrium if and only if
\begin{equation}\label{eq:Nash}
    u_i(a_i^*,a^*_{-i})=\max_{a_i\in A_i} u_i(a_i,a^*_{-i}), ~\forall i\in N.
\end{equation}
Thus, if player $i$ finds that
the other players select $a^*_{-i}$, $i$'s payoff is maximized by choosing action $a^*_i$ (perhaps not uniquely).




\subsubsection{Stackelberg Games}
An extension of Normal form games is the \emph{Stackelberg game}, where one player may first see the other's action before making their own move. It is a sequential-move, one-shot, finite-action, two-player game.
This game is commonly applied to model that the defender first commits to a configuration, then the attacker may observe and best respond. For example, the defender chooses their 
IDS placements in a network first, and the attacker observes this distribution and plans their attack accordingly \cite{carter2014game,sengupta2018moving,chowdhary2018markov}.
This game has also been applied to model the situation where the defender must decide on how to allocate resources between patching known vulnerabilities and deploying honeypots, after which the attacker explores and exploits the network 
\cite{jajodia2018share}.
The Stackelberg game can be further generalized into an \emph{Extensive Form game},
where players make choices one at a time, and each can react to the previous choices made.
This game may permit any number of players and actions per play, with the only restriction that the order of play must be predetermined.
This game has been applied to study Insider Threats 
\cite{feng2015stealthy}, where the defender first selects their defensive configuration, then an insider makes an offer of information to the attacker, who then either accepts or declines it before selecting an attack strategy.

Denote by $N=\{L,F\}$ the ``leader" and ``follower" in Stackelberg games, respectively.
The solution concept is an adapted version of the Nash equilibrium, called {\em Stackelberg equilibrium}, which is the action profile $a^*$ that solves the following
\begin{equation}\label{eq:Stackelberg}
\max_{a_L\in A_i} \min_{a_F\in \br_A(a_L) } u_L(a_L,a_F),
\end{equation}
where the follower $F$'s best response function is given by $\br_A(a_L)=\argmax_{a_F\in A_F}U_F(a_F,a_L)$.
Intuitively, Stackelberg equilibrium captures a similar situation as Nash with the addition of the decision-making order.
At equilibrium the follower cannot profitably deviate given the leader's action, and if the leader deviates there exists a best response by the follower such that the leader's utility will decrease.
We discuss the computation of the Stackelberg Equilibrium in Section~\ref{sec:LP}.

In cybersecurity, the Stackelberg equilibrium is proposed as a computable solution to the cybersecurity situation where the attacker can observe the defensive configuration \cite{sengupta2017game,sengupta2018moving,chowdhary2018markov,umsonst2018game}.
Although it is often impossible to predict the attacker's action, the Stackelberg equilibrium has the claim that a rational attacker will play their own equilibrium action (in response to the defender's equilibrium action).
Thus, the utility received in the computed Stackelberg equilibrium can act as a meaningful predictor of the utility that will be realized during the actual deployment.
One remarkable feature of these games is that they manifest Kerckhoff's Principle in cybersecurity as the attacker is given information about the defense posture before deciding how to wage attacks; the Principle was introduced in the context of cryptography 
\cite{GoldreichBookVol1}.

\subsubsection{Stochastic Games}

These models seek to develop disciplined defense strategies against stealthy or undetected attackers. This game is named because there is an underlying state or environment that changes randomly, possibly as a function of the play of players. It defines a (usually finite) state space $S$ such that $s\in S$ denotes a specific cybersecurity state, and some randomized transition rule $P(s',s,a)=Pr(S_{t+1}=s'\mid S_t=s, A_t=a)$ that defines how the game moves to state $s'$ given the previous state $s$ and actions $a$ taken by players.
It has been applied to model MTD \cite{lei2017optimal,tan2019optimal,gothawal2020anomaly,horak2019optimizing,carter2014game}, where $S$ defines the state of a network (possibly including a history of recent network configurations).
The players select $\sigma_i(s)\in \Delta(A_i)$ which is, as defined above, a state-dependent distribution over their actions for each state $s$.
These strategies induce a Markov chain that can be analyzed via utility function $u_i(s,a)$, which typically depends on both the state and the player's actions.

\begin{figure}[!htbp]
    \centering
\includegraphics[scale=0.075]{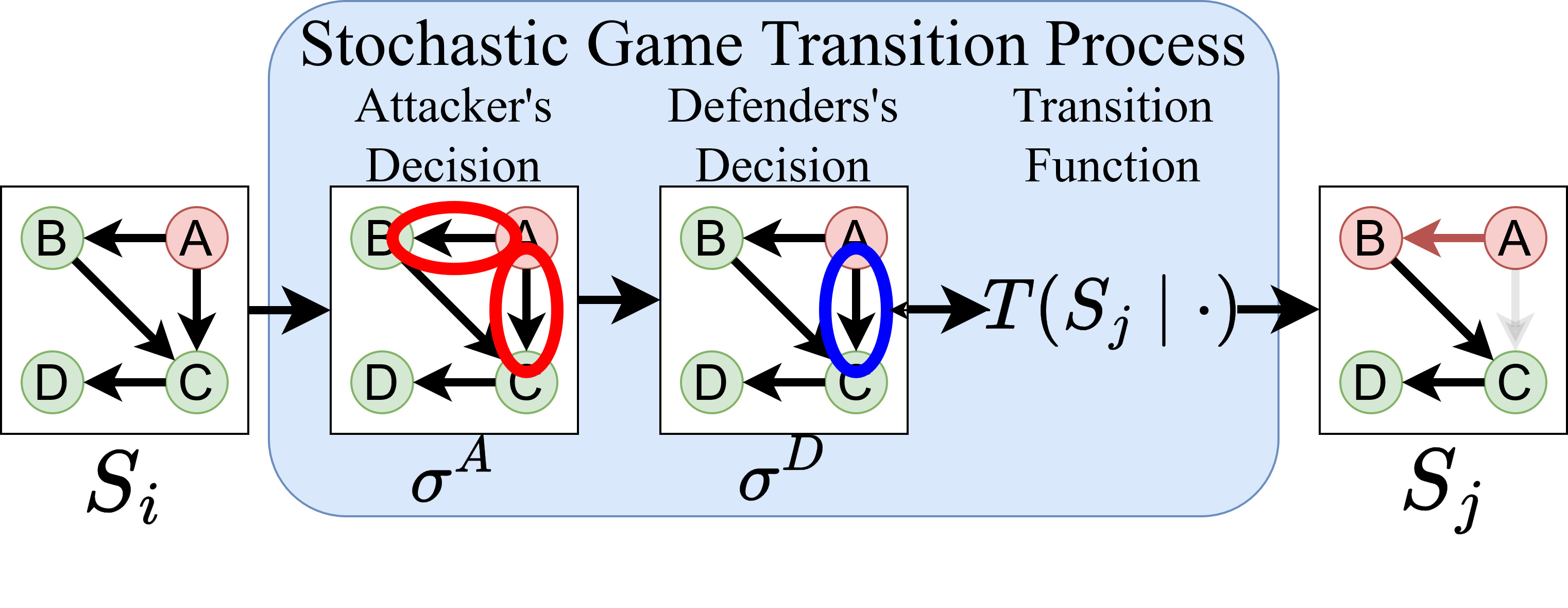}
    \caption{Stochastic Game model of a network security MTD problem. }
    \label{fig:stochastic game}
\end{figure}

To illustrate how a stochastic game can model MTD, we consider the network example presented in \cite{lei2017optimal}.
Suppose a defender possesses a computer network, visualized in Figure~\ref{fig:stochastic game} as computers $A,B,C,D$ with arrows that define the permitted connections between computers.
Let green indicate that a computer is controlled by the defender and red indicate the attacker has compromised it.
Let the network as a whole have a finite number of security states, denoted by $S=\{S_1,S_2,\dots,S_k\}$.
These states represent the network firewalls, topology, IDS configurations, as well as any other defensive tools that a defender might have.
Additionally, suppose the attacker has a list of vulnerabilities that allow them to infect additional nodes, escalate their privileges, or deliver a payload.
The model then proceeds with the attacker choosing a vulnerability that the defender observes and then chooses defensive tools.
The attacker's and defender's action then jointly alters the security state $S_i$ of the network.
The changes that these actions have on the original state $S_i$ are encoded by the probabilistic function $T$, which allows for randomized considerations such as the fact that actions of both the defender and attacker can have a chance of being unsuccessful.
The reward for both parties then depends on the state $S_j$ achieved and the operational costs associated with their action.
The attacker seeks a probabilistic policy $\sigma^A$ that assigns a distribution to their attacks conditioned on the current network state $S_i$, while the defender seeks policy $\sigma^D$ that assigns a distribution to their available defensive actions given the attacker's action and the state.

\begin{figure}[!htbp]
    \centering
    \includegraphics[scale=0.12]{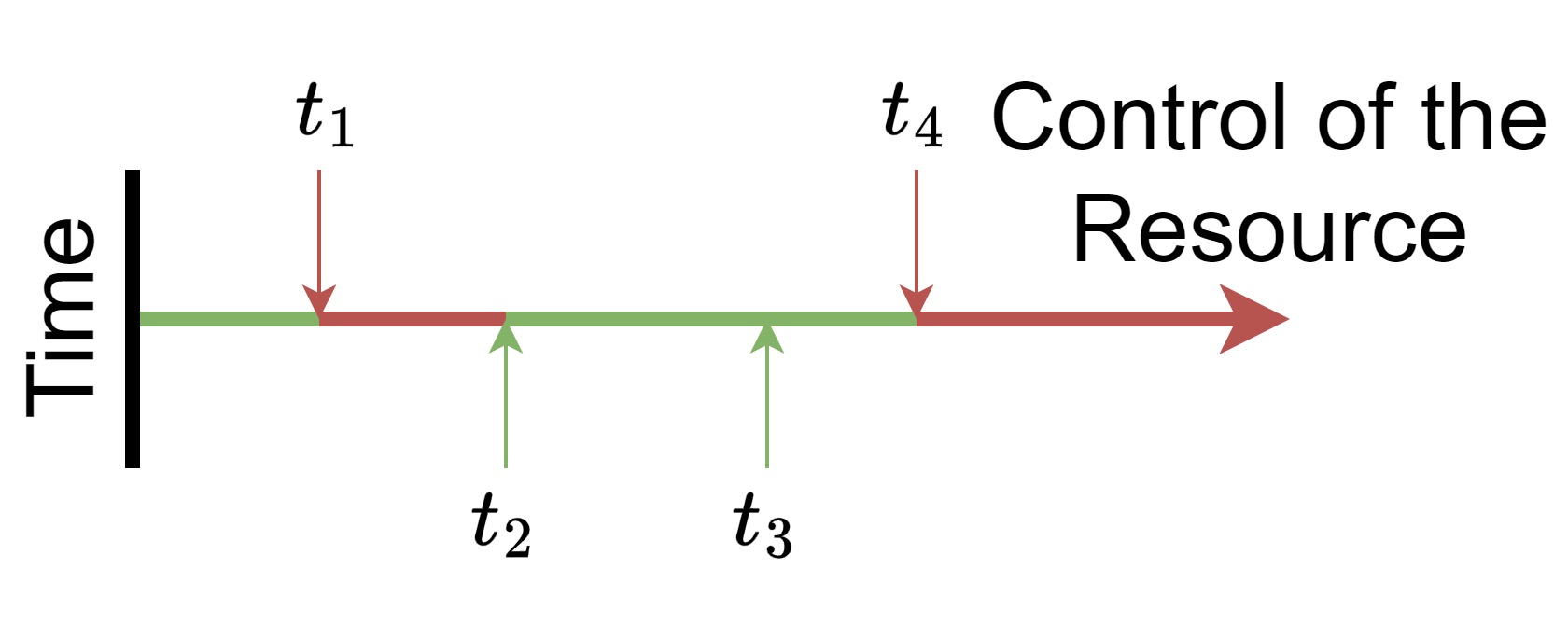}
    \caption{The FlipIt Game Model. The red color indicates the attacker controls the resource and the green indicates the defender.}
    \label{fig:FlipIt}
\end{figure}

A special class of stochastic games that was developed specifically for the APT problem is \emph{FlipIt}.
The original FlipIt model \cite{van2013flipit} focused on the core assumption that defenders in APT scenarios often make a decision with little knowledge.
Specifically, the defender has a resource (e.g., computers) that both the attacker and defender can take control of at any point in time.
While neither player knows who has current control of the system, both players attempt to maximize the time they control the resource, with the consideration that every action they take has a cost.
This is depicted in Figure~\ref{fig:FlipIt}, where the color represents control of the resource, and each of $t_1,t_2,t_3,t_4$ represents one party taking control.
Intuitively, each party wants to maximize their time in control of the resource, however, it is assumed that taking control incurs some cost (e.g., reinstalling the operating system takes time).
Thus, each side must balance the cost of taking control with the benefit of having control.
The complications of this are seen at $t_3$, where the defender takes control of the resource, despite unknowingly already having control of the resource.
Thus, this action had no benefit and represents the danger of taking action too frequently and incurring unneeded cost.
Conversely, the danger of acting too infrequently is the risk of little time in control of the resource.
Thus, the APT model seeks to find a policy that balances the cost of dilemmas.

The goal of FlipIt is to model the stealthy nature of APTs; as such, neither player knows who has control of the resource and either player can seize control of it at any given time.
The pitfall of this model is that the assumption that the attacker does not know which nodes have been compromised is largely flawed, and further the defender having no knowledge of the security of their system is also unrealistic.
A defender often has some (possibly noisy) information about the security state of their systems based on IDS behavior and logs.
This model has seen many extensions, including multiple resources \cite{laszka2014flipthem}, informed attackers \cite{zhang2014stealthy,zhang2015game}, partially compromised resources \cite{farhang2016flipleakage}, insider threats \cite{feng2015stealthy}, Prospect Theory \cite{xiao2017cloud}, and others \cite{pawlick2015flip}.


The solution concept is a Nash-based condition that accounts for the time horizon, namely an equilibrium where no player can unilaterally deviate their policy with respect to a discounted or average utility in an infinite time horizon or the sum of a finite time horizon.


\subsubsection{Bayesian Games}\label{ssec:bayesian}

In cybersecurity, attackers and defenders often have limited knowledge of each other and may learn each other's intentions and capabilities over time and adjust their behavior accordingly.
Bayesian games fit this situation naturally.
These games are constructed like standard games with players, actions, and utilities, except that each player has a private {\em type} $\theta_i\in \Theta_i$ which encodes some private information about their utility function,
capabilities, and identities  (unknown to the other players)
\cite{wang2017strategic,tian2019defense,diamantoulakis2020game,du2019sdn}.
In the simplest formulation, these games are single-stage games where each player $i$'s type is sampled from a known joint distribution, and each player $i$ can use knowledge of this distribution to form beliefs about the types 
of all other players.
Players can use Bayesian inference to compute their optimal behaviors in expectation over their beliefs.
The solution concept is {\em Bayes-Nash equilibrium}, which builds on the Nash condition given by Eq.\eqref{eq:Nash} while incorporating the player's beliefs.



One special case of Bayesian games is the \emph{multistage Bayesian game}, where the attacker and defender learn about each other through repeated interactions.
\cite{basak2019identifying,liu2006bayesian,la2016deceptive,huang2020dynamic,niazi2019bayesian,bouhaddi2018efficient,zhu2018multi,huang2019adaptive}.
The most common solution concept applied to multistage Bayesian games is the \emph{perfect Bayesian equilibrium}, which is defined under two conditions: \emph{sequential rationality} and \emph{belief consistency}.
Sequential rationality is analogous to the Nash condition given by~\eqref{eq:Nash}, 
and ensures that no player can profitably unilaterally deviate their strategy at equilibrium.
Belief consistency requires that all players update their beliefs about other players using Bayes' Theorem.


Another special case of Bayesian games is the \emph{Signaling game}, which also emphasizes the asymmetric nature of information in cybersecurity.
In Signaling games, the leader gets to send a signal to the follower, who must decide how much to trust the signal based on the prior beliefs.
Signaling games have been applied to study MTD \cite{feng2017signaling}, where the defender can choose a policy and signal a configuration independently of the configuration that is actually selected.
In \cite{aydeger2021strategic}, the problem of DDoS attacks is considered, where the attacker or spammer is regarded as the signaler and the defender must discern if the user is legitimate or a DDoS attacker.
The solution concept is also the perfect Bayesian equilibrium.

\subsubsection{Differential Games}

Differential games can handle the situation where players make decisions on a continuous timescale, which is not ``natural'' in cybersecurity where time is inherently discrete. 
Nevertheless, these models offer an alternative, which possibly permits analyses that would not be feasible otherwise.
These games have been applied to study CTI sharing \cite{hausken2007information,gao2014game,gao2016differential}, where firms (or enterprises) can choose their rate of sharing CTI.
They have seen application in the study of APTs \cite{hu2015dynamic,yang2018effective,li2017differential,yang2018risk} and cyber-physical systems \cite{fotiadis2020constrained}.

To see how differential games function in terms of cybersecurity, we give an example of CTI sharing \cite{tosh2015cyber}.
This situation is which are well suited to situations that can be modeled by dynamical systems (i.e., inputs influence the derivative of a state function).
In this work, they consider the situation where $n$ firms may choose to collaborate with a trusted third-party CTI aggregation and analysis center.
Each firm $i$ may choose how much money to invest in their cybersecurity in the way of finding vulnerabilities in their network.
Then, the firms select how much of the discovered information to share about a given vulnerability to the third party, who then aggregates all of the information and returns it to all $n$ firms.
In addition to incurring the cost of the investment, firms also incur a reputation loss increasing with the amount of intelligence shared.
This model is known as a Differential game, as these investment and sharing decisions are updated continuously as the situation evolves and defines the derivative of each firm's utility function.
Thus, an investment and sharing schedule for each firm defines a dynamical system.

The solution concept is typically a continuous version of the Nash equilibrium, termed \emph{saddle points}. In the context of CTI sharing, the solution concept can be leveraged to compute 
``market share'' of each firm resulting from the game, which can be interpreted as
the cybersecurity gain benefiting from participating in CTI sharing.

\subsubsection{Evolutionary Games}
\label{sec:evolutionary}
{\em Evolutionary games}  were originally proposed to study ecological systems, where each player is a species of ``animal'' that inhabits a place with finite resources. These models aim to predict what mix of species will occur in the long run based on population interactions (or actions) such as predation.
These games have been adapted to study \cite{shi2021research,
tian2021honeypot,
abass2017evolutionary,
tosh2015evolutionary,
tosh2018establishing}:
(i) honeypot deployment \cite{shi2021research,tian2021honeypot}, where the defender decides whether to deploy a honeypot and the attacker decides whether to attack; 
(ii) APTs 
\cite{abass2017evolutionary}, where defender and attacker 
determine how often they take control of the system in question; and 
(iii) CTI sharing \cite{tosh2015evolutionary,tosh2018establishing}, where a population of firms decide on whether to share their CTI with each other.

The associated solution concept is known as the \emph{Evolutionary Stable Strategy (ESS)}, which is a refinement of the Nash equilibrium with added {\em stability} properties.
In addition to the Nash condition given by Eq.\eqref{eq:Nash}, an ESS must be resistant to incursions of other strategies at equilibrium, owing to 
that many random mutations of species occur over time. 

\subsubsection{Coalitional Games}

\emph{Coalitional games}, also known as {\em cooperative games}, focus on cooperation among the players.
This has natural interpretations in cybersecurity because defenders (e.g., firms or enterprises) may cooperate with each other in cyber defense.
In these models, a set of defenders $N$ mutually agree on a coalition $C\subseteq N$ that benefits them in contexts such as CTI sharing \cite{vakilinia2017coalitional,vakilinia2019fair}, communal cybersecurity insurance \cite{vakilinia2018coalitional},
and IDS \cite{guo2018incentive}.
For example, IDSes that are deployed at various places of cyberspace may cooperate with each other to increase their confidence in flagging an event as attack \cite{guo2018incentive}. 
The solution concept is a stable coalition that members of the coalition (e.g., IDSes) find their best interest to continue to participate in the coalition.

\subsubsection{Gestalt Games}

\emph{Gestalt games} model multistage attacks
\cite{pawlick2015flip}, \cite{zhu2018multi}.
These games chain together multiple game-theoretic models into {\em stages}, where the outcome of previous stages impact future ones.
One example is to model the security of a cloud server as a FlipIt game, where the cloud sends a signal to an autonomous player (i.e. a remote drone) that must decide whether to trust the signal \cite{pawlick2015flip}.
Another example is to model APTs  where each stage of the game represents a stage in the cyber-kill chain as follows: 
the game begins with a Normal form game modeling an MTD scenario, whose result impacts the attackers' effectiveness in later stages \cite{zhu2018multi}.

The solution concept, \emph{Gestalt equilibrium}, consists of the solution concepts of the component games. 
For example, a Gestalt equilibrium could consist of a Nash equilibrium, 
followed by a Bayes-Nash equilibrium, where the results of the earlier equilibrium affect the parameters of subsequent games.

\section{Efficacy and Practicality of Proposed Models}
\label{sec:metrics}

Ultimately, the goal of game-theoretic modeling for cybersecurity is to produce useful solutions that cybersecurity practitioners can implement and utilize.
To evaluate how well the literature is achieving this goal, we evaluate two orthogonal directions: \emph{Efficacy of Solutions} and \emph{Practicality of Solutions}.
The first, Efficacy of Solutions, seeks to assess how useful the solution will be based on how well the model captures the first principles, or the reality, of the underlying cyber interaction.
To assess this question, we propose a list of ideal model attributes for each application based on the interaction.
Then, we analyze how well the assumptions of the typical modeling approaches address our list of first-principle primitives.
The difference between these two is the deficit that future works should strive to remedy.

In the second direction, Practicality of Solutions, we seek to assess whether the literature solutions can be obtained in practical settings.
Particularly, we evaluate the informational aspect of whether or not practitioners could access all of the necessary information to compute the prescribed equilibrium.
To investigate this, we created the Player Knowledge section of Table~\ref{tab:assumptions} (i.e. Section \ref{sec:player knowledge}), which summarizes the information assumptions needed to compute the prescribed equilibrium.
For conciseness, we summarize this information by game-theoretic model in Table~\ref{tab:knowledge by model}, as well as the popularity of each model in each application.
This section makes the assumption that the solution concept found in the work (usually an equilibrium) is a prescription for practitioners to implement in the relevant cybersecurity situation.
Therefore, 
Section \ref{sec:player knowledge} and Table~\ref{tab:knowledge by model} are lists of the information required for a cyber defender to gain useful insight from the model.

However, because learning of the various parameters also depends on the cyber situation, we also investigate the practicality of knowing each piece of assumed knowledge with respect to its problem domain.

\begin{table}[!htbp]
\begin{tabular}{|l||l|l|l|l||l|l|l||l|l|l||l|l|l|}
\hline
             &IDS&APT&MTD&CTI& $a$   & $s$   & $h$   & $A$   & $N$   & $U$   & $\mathcal{A}$ & $\mathcal{N}$ & $\mathcal{U}$ \\ \hline
Normal Form  &6&2&\textbf{8}&2& \cN &  &  & \cY & \cY & \cY & \multirow{9}{*}{\small\rot{Unmodeled}}           &\multirow{9}{*}{\small\rot{Unmodeled}}            & \multirow{4}{*}{\small\rot{Unmodeled}}           \\ \cline{1-11}
Stackelberg  &1&1&\textbf{3}&0& \cN &  &  & \cY & \cY & \cY & & &  \\ \cline{1-11}
Stochastic   &2&1&\textbf{9}&0& \cN & \cY &  & \cY & \cY & \cY & & &  \\ \cline{1-11}
FlipIt       &0&\textbf{10}&1&0& \cN & \cN &  & \cY & \cY & \cY & & &  \\ \cline{1-11}\cline{14-14}
Bayesian     &3&\textbf{6}&3&0& \cN &  & \cY & \cY & \cY & \cN & & & \cY$^*$ \\ \cline{1-11}\cline{14-14}
Differential &1&4&1&\textbf{5}& \cN & \cY &  & \cY & \cY & \cY & & &  \multirow{4}{*}{\small\rot{Unmodeled}}\\ \cline{1-11}
Evolutionary &1&0&\textbf{2}&\textbf{2}& \cN &  &  & \cY & \cY & \cY & & &  \\ \cline{1-11}
Coalitional  &1&0&0&\textbf{2}& \cN &  &  & \cY & \cY & \cY & & &  \\ \cline{1-11}
Gestalt      &0&\textbf{2}&0&0& \cN &  &  & \cY & \cY & \cY & & &  \\ \hline
\end{tabular}
\caption{Popularity and typical knowledge assumed by model type. The first grouping of columns counts the number of models of each type surveyed by application area (Bold indicates the most popular application for each model), and the subsequent columns showcase the typical knowledge required to compute its associated equilibrium.
A green checkmark (\cY) indicates that the cyber defender needs to know the corresponding model parameter in order to calculate the relevant solution concept, conversely red X (\cN) indicates they do not need to know that parameter, and a blank space indication there is no notion of that parameter in the model. *Bayesian games most commonly consider uncertainty of the action set, but occasionally consider other forms of uncertainty.  \label{tab:knowledge by model}}
\end{table}

To justify our approach of analyzing the popular game-theoretic models, we observe that most approaches use the same few models.
Often, game-theoretic modeling in cybersecurity is done by applying an off-the-shelf game-theoretic model and describing its parameters in terms of cybersecurity. 
Although friendly for (typically equilibrium) analysis, this creates strong homogeneity of applied models and leaves important aspects of cyber-interactions unmodeled.
To see evidence of this homogeneity, we aggregate some statistics from Table~\ref{tab:assumptions}:
the top five most popular models (Normal, Bayesian, Differential, FlipIt and Stochastic) make up 74\% of all models used, 79\% of all models use the Nash equilibrium and adapted variants (Nash, Bayes-Nash, Stackelberg) as a solution concept, and 76\% of models only consider a two-player interaction.

Before beginning our analysis for each application, we highlight two universal suggestions that can aid in both directions.
The first suggestion pertains to the way the authors state their assumptions.
Traditionally in game theory, games are stated by the tuple $(N,A,U)$ leaving other elements of the interaction implicitly defined in text. 
This is problematic because equilibrium typically requires knowledge of all aspects of the interaction, and therefore finding out the knowledge needed to compute the equilibrium can require significant game-theoretic expertise.


To bridge this deficit, we advocate using our framework to fully and explicitly state a model, including its assumed model structure $(\mathcal{N},\mathcal{A},\mathcal{U},R,T)$ and player knowledge $$K^c=\{\mathcal{N},\mathcal{A},\mathcal{U},N,A,U,a,T,R,S,s,h\}.$$
Explicitly stating the player knowledge required to utilize the proposed solution concept simplifies evaluating the practicality of proposed solutions.
Specifically, increasing the practicality of game-theoretic solutions (by reducing the information needed to produce a solution) is a significant contribution to game-theoretic cybersecurity.
Thus, formally stating these assumptions will make understanding the state-of-the-art easier and contributions more clear.

Secondly, we highlight the need to model the uncertainty of the adversary in cybersecurity.
Unfortunately, we find that uncertainty over the set of adversaries, their actions or capabilities, and utility functions are underexplored.
Particularly, existing models typically assume that there is one possible set of players, one set of actions, and one set of utility functions (with the exception that Bayesian games allow uncertainty, usually over the utility function).
This largely removes notions of uncertainty common to many adversarial cybersecurity problems, such as the identity, intentions, and capabilities of cyber attackers.
Universal to all applications, we make the following three recommendations:
\begin{itemize}
    \item \emph{Uncertainty of the Player Set.} We advocate for models to use the notion of $\mathcal{N}=\{N_1,N_2,\dots\}$ to capture the situation where many sets of adversaries are possible and the defender must discover which $N\in\mathcal{N}$ is correct.
    \item \emph{Uncertainty of the Action Set.} We advocate for models to use the notion of $\mathcal{A}=\{A_1,A_2,\dots\}$ to capture the situation where many action sets may be possible and the defender must learn the capabilities $A\in\mathcal{A}$ of their adversaries through interaction.
    \item \emph{Uncertainty of the Utility Functions.} We advocate for models to use the notion of $\mathcal{U}=\{U_1,U_2,\dots\}$ to capture the situation where the adversary's objectives are unknown and must be learned through interaction.
\end{itemize}
We now evaluate the efficacy of each application given its needs.
Additionally, we highlight which of the above three notions of uncertainty are most necessary in each area.

\subsection{Efficacy and Practicality of IDS Optimization Models}
The primary question being addressed by IDS Optimization models pertains to the trade-off between the cost of over-flagging versus the benefit of increased security.
Thus, it can be assumed safely that the adversaries' actions are known and observable since they make up the content that an IDS must discern as malicious.
Beginning with efficacy, we have developed the following list of traits that an IDS optimization model should have:
\begin{enumerate}
    \item Provide guidance on how an IDS should flag; 
    \item leverage the computer network;
    \item formalize the model's relationship with time; and
    \item incorporate uncertainty over the player set and the utility functions.
\end{enumerate}


IDS optimization has primarily been studied using Normal form and Stackelberg games, as discussed in the example in the previous section.
We now evaluate this approach with respect to the list of desired objectives.
First, we acknowledge that, given full information, both Normal form and Stackelberg games do provide mixed strategies that inform how aggressive an IDS should be.
However, without any real source of data, the numbers used in the models are unitless and arbitrary.
For example, in the example given in Section~\ref{ssec:normal form game}, a false alarm is 20\% as bad as a breach, which is not particularly well justified with respect to a real world scenario.
Therefore, the application of cybersecurity metrics, particularly studying the impact of IDS false alarms and breaches, would lead to better modeling.
Secondly, IDSes typically do not exist in isolation, and the state of the network the IDS exists within can help inform how it should flag items.
Thirdly, the Normal-form game and Stackelberg game matrix are intrinsically static with respect to changing parameters.
Thus, how and when to recalibrate them is itself a nontrivial question that must be investigated before adopting these models.

We highlight that IDSes may experience many interactions from many attackers.
This implies that it is necessary to distinguish between multiple attackers and to consider the problem of threat attribution.
For this purpose, we advocate for using the notion of the space of player sets $\mathcal{N}$ to specify, for example, the situation where there may be one, two, or three attackers via $\mathcal{N}=\{\{1,2\},\{1,2,3\},\{1,2,3,4\}\}$, while noting that the defender is also a player.
This formalizes the notion that there may be multiple attackers in the model, and which player set $N\in\mathcal{N}$ is not assumed knowledge but something the defender must learn from interactions with adversaries.
Finally, different attackers have different objectives and it may be pertinent to prioritize flagging suspected activity of one attacker over others.

We evaluate the practicality of solutions derived from Normal form and Stackelberg games.
These are, as discussed in Section~\ref{sec:models}, these solutions are the Nash equilibrium and the Stackelberg equilibrium.
From Table~\ref{tab:knowledge by model}, we can see that both Normal form and Stackelberg games assume knowledge of $A,N,U$.
Second, assuming that the IDS knows that there is a single attacker is another significant assumption. 
Beginning with the utility function (represented by the numbers within the matrix), knowledge of this implies that the defender knows their preferences and the adversary's evaluation of each possible scenario.
Specifically, the defender knows the cost of a false alarm, the expected cost of the breach, and the attacker's incurred cost of attacking and perceived benefit of a flagged attacked.
The first two that pertain to the defender's own parameters could be measured, but measuring the second two that pertain to the adversary's preferences would be much more difficult.
To remedy this, there are several directions.
First, the IDS could attempt to estimate the adversary's parameters from simulated or real data.
Second, the robustness of the equilibrium could be studied to account for errors in the previous estimator.

In summary, we find:
\begin{itemize}
    \item\textbf{The Good.} IDS optimization models provide guidance when to flag given full information, and it is reasonable to assume that an IDS can learn parameters regarding itself.
    \item \textbf{The Bad.} Existing approaches fail on the latter three criteria: leveraging the computer network, formalizing the model's relationship with time, and incorporating uncertainty.
    \item \textbf{The Ugly.} To model the strategic interaction with an adversary, it is necessary to have at least an estimate of their utility function.
    No approach addresses the question of learning this or considers robustness of bad estimates.
    Finally, the current models have no capacity to incorporate learning this utility function over time.
\end{itemize}

\subsection{Efficacy and Practicality of APT Defense Models}
APT Modeling is a unique challenge as it emphasizes sophisticated and stealthy attacks.
From a modeling point of view, this makes it difficult to devise an information model for the defender.
This is because the defender often cannot observe the direct actions of the attacker, but rather noisy signals from an array of sources such as an IDS.
Second, the multistage nature of APT's all have fundamentally different interactions, requiring models that can accurately capture each stage. 
An ideal model for tackling APT would have four characteristics:
\begin{enumerate}
    \item incorporate realistic information measurement models, such as noisy IDS alerts and stealthy attacks; 
    \item incorporate chained threats and dependencies into outcome evaluation,
    that is the network structure of computers and their relative importance must be considered;
    \item incorporate distinct stages of an APT from
a framework like MITRE ATT\&CK such as reconnaissance, privilege escalation, and lateral movement; and
    \item incorporate uncertainty and learning of the player set, action set, and utility functions.
We stress that all three notions of uncertainty exist in an APT setting.
\end{enumerate}

APTs have been primarily modeled using FlipIt \cite{van2013flipit}, Bayesian games, and Gestalt Games.
FlipIt has the advantage that it was designed specifically for this application, and accordingly models situations where the defender does not know the state of their own system.
The goal of FlipIt is to model the stealthy nature of APTs; as such, neither player knows who has control of the resource and either player can seize control of it at any given time.
With regard to our proposals for the APT modeling space, we find that FlipIt succeeds in modeling stealthy attacks although the zero information assumption is perhaps too strict.
Ideally, future FlipIt models could incorporate sources of noisy information such as IDSes, suspicious activity, or evidence of a cyber breach (such as erratic control behavior in the context of StuxNet).
However, no FlipIt-based approach succeeds in any other of the three criteria.

In contrast, the main advantage of Bayesian games is that they are the only model that enables the defender to learn about the adversary over time, such as their strength or intention. 
Their weakness is that they do not capture the stealthy aspect of APTs as Bayesian games require the defender to observe previous attacker actions to update their belief.
Typically, Bayesian games consider adversaries to have a type that determines their utility function \cite{huang2018analysis,huang2019adaptive,huang2020dynamic}, but other properties have also been investigated, such as their starting point in a network \cite{rass2019cut} and their set of vulnerabilities \cite{basak2019identifying}.
Thus, Bayesian games have been successfully applied to consider chained dependencies and uncertainty over the action set and utility functions.
Although Bayesian games do implement uncertainty over both the utility and action sets, they do it in a fairly limited manner.
Models typically consider the private type to be a unit-less continuous quantity representing adversary strength \cite{huang2018analysis,huang2019adaptive}, or a small finite set representing an attribute of the adversary \cite{huang2020dynamic}.
This leaves only realistic information models and incorporating distinct APT stages completely unaddressed.
One approach to increasing the Modeling Efficacy would be to base the private Bayesian types on real cybersecurity quantities, such as considering a system that has $x$ vulnerabilities and the adversary has access to some portion of them.
Then, which subset of capabilities the adversary has must be learned over time based on their observed actions.
Further measurement or justification is needed to validate that a model with so few types is the correct approach.
In summary, Bayesian games succeed at objectives 2 and 4, but lack realistic information models and distinct stages of an APT.

This issue of not considering distinct stages of an APT is specifically remedied by the proposed Gestalt game, which consists of arbitrary combinations of elementary games played sequentially such that the outcome of a game affects the parameters of the next game.
This approach, as proposed in \cite{zhu2018multi}, enables the distinct stages of an APT to be studied although it has yet to incorporate learning about the adversary's intentions during play, using a dependency-based model or accommodating stealthy attacks.
Thus, across all three models proposed for the APT domain all four of the criteria have been addressed.
However, no single comprehensive model exists to handle all four objectives in a single model.
Therefore, future approaches should seek to cross-pollinate ideas and analysis techniques to achieve more effective modeling in this area.

Next, we evaluate the practicality of computing the solutions from the proposed FlipIt, Bayesian games, and Gestalt games. 
We first evaluate the practicality of knowing $A,N$ which is assumed in all three games (from Table~\ref{tab:knowledge by model}).
With respect to knowing the action set, the premise of stealthy attack is fundamentally incomparable with knowledge of the action set.
Indeed, in any APT scenario there is significant uncertainty of the adversary's capabilities, which remains an under-investigated component with respect to game-theoretic models.
Second, models universally know the player set and is usually assumed to be a single adversary.
However, it is necessary to determine that there is a single adversary or how to attribute individual attacks to an adversary.
Therefore, future APT models should consider either a real mechanism for cyber attribution or novel solution concepts that do not depend on precise knowledge of the player set.

Moving to the utility functions, both the FlipIt and Gestalt games depend on complete knowledge of all players' utility functions.
As mentioned above, this implies that the defender is able to measure the cost of their defensive actions, the cost of intercepted breaches, and unmitigated breaches.
It would be ideal if the approaches directly include these measurements as part of the model, because current approaches use arbitrary and unit-less values for the utility function.
More troubling, though, they depend on knowing adversaries' costs for attacking and gain if they succeed.
Although Bayesian games allow for uncertainty over a set of utility functions, it is worth noting that typically those models only consider two possible utility functions.
As it stands, assuming that only two utility functions, or some significant knowledge of the structure of utility function is too strong for cybersecurity application.
Therefore, it is necessary that future work develops methods to learn the attacker's utility functions.
Two approaches are possible: first, there is some data surrounding the cost of exploits \cite{meakins2019zero} that could be used to develop a rigorous methodology to estimate attacking costs.
A second approach is to develop novel solution concepts that can handle greater uncertainty of the attacker's utility function.
Ideally, both approaches can be used together to generate rigorous and measurable solutions for cyber-defenders.

Finally, Bayesian games additionally assume that the history of play is observable.
This is necessary for the defender to attempt to determine the adversary's private type is.
In some formulations, this is justified, such as \cite{huang2020dynamic} where the defender attempts to ascertain a malicious user from a legitimate one.
The assumption is well-justified in the case that the defender can observe all of the attacker's actions, however, not in scenarios where the attacker employs stealth.
Therefore, future models should include realistic noisy measurements of models of past actions such as IDS readings, manual file forensic analysis, or suspicious network activity.

In summary, we find:
\begin{itemize}
    \item \textbf{The Good.} The first three objectives have been satisfied by various models, and the fourth partially satisfied by Bayesian Games.
    \item \textbf{The Bad.} No single model simultaneously achieves all four objectives, making it difficult to piece together guidance from multiple models.
    Additionally, uncertainty of the player set and threat attribution require a much more rigorous characterization.
    \item \textbf{The Ugly.}
    The notion of stealthy attack and assuming one knows the adversary's action set (assumed by all models) are fundamentally incompatible.
    Future models should seek to characterize the dynamics between a resourceful attacker with uncertain capabilities and a defender leveraging noisy measurements.
\end{itemize}

\subsection{Efficacy and Practicality of MTD  Models}
In the MTD domain, game-theoretic models attempt to derive switching policies that balance increased security with costs associated with switching.
MTD is a strong candidate for game-theoretic modeling due to its automatic nature.
In addition to creating a switching policy, we outline several attributes that an ideal MTD model would have:
\begin{enumerate}
    \item optimize {\em when}, {\em where}, and {\em how} to switch between configurations simultaneously;
    \item include technical measurements in parameter calculation (i.e. what is the cost of switching);
    \item recalculate parameters in a dynamic manner;
    \item incorporate uncertainty and learning of the player set, action set, and utility functions.
\end{enumerate}

MTD is modeled primarily using Normal Form, Stackelberg/Extensive Form, and Stochastic games.
Normal and Stackelberg/Extensive form games explicitly model the distribution over possible configurations to be the defender's decision variable.
In the normal form game, the defender's configuration can be updated strategically with respect to the attacker's choices, while in the Stackelberg \cite{sengupta2018moving} they must commit to a distribution that the attacker can observe and respond to.
Stochastic games enable analysis of higher-fidelity policies, as the defender does not choose a single probabilistic strategy but instead an individual strategy for each security state or defensive configuration.
All of these games offer only a partial solution to the classical ``where, when, and how'' of MTD switching.
In the first case of normal form and Stackelberg games, a distribution provides criteria of a MTD policy (i.e. how often should the system be in each state), but does not actually specify at what rate the switching should occur or how the transitions from states should occur.
The Stochastic game does slightly better in answering the how, as its policy is a function of the present state, informing how the defender should transition from one state to the next.
Additionally, this model incorporates the notion of switching costs, which cannot be directly accounted for in normal form/Stackelberg games.

With regard to the second objective, Stackelberg games \cite{sengupta2017game,sengupta2018moving} have utilized the Common Vulnerability Scoring System (CVSS) which uses experts that categorize various features of vulnerabilities into numerical scores that quantify the severity and ease of exploitability of the exploit.
This addresses using real world data to calculate the parameters for the problem.
Unfortunately, neither method addresses any of the other two criteria.
Parameter recalculation is necessary in MTD defense, or a novel model is proposed that accounts for changing parameters over time.
Finally, in many application areas of MTD, such as honeypot deployment or IDS deployment, the defender gains information upon their defensive countermeasure interacting with the adversary.
This provides information about the attacker's capability, objectives, and identity that could be leveraged to form a better defense.

Moving on to evaluating the practicality of these solutions, we now evaluate the practicality of the assumptions associated with Stackelberg games and Stochastic games in the context of MTD.
From Table~\ref{tab:knowledge by model}, it can be seen that all models require knowledge of the action set, player set and utility functions.
In normal form and in Stackelberg games, the defender action set is the set of configurations and the attacker's actions are a set of vulnerabilities \cite{lei2018incomplete,tan2019optimal}. 
Following this intuition, it is fairly reasonable that the defender knows the full set of configurations they have access to and the attacker knows the full set of exploits they can do.
However, it is less clear that knowledge of the full set of the other's capabilities is justified.
If the attacker is restricted to using disclosed vulnerabilities, considered in \cite{sengupta2017game,sengupta2018moving}, then this assumption is justified as commercial software scanners are capable of measuring this set.
Additionally, more application-specific analysis is required to determine whether the assumption that the attacker has knowledge of all possible configurations is justified.
Secondly, the player set is assumed to be known.
This raises the question of attack attribution, that is, given a series of observed cyber attacks in a network, how does one determine how many unique attackers are present?
No work in this domain investigates this question or justifies it with respect to the specific proposed application of MTD.
Therefore, future works should justify either why it is acceptable to assume knowledge of the player set or provide justification for how it will be measured.

Finally, computation of equilibrium models requires knowledge of both the defender's and the adversary's utility functions.
From the defender's point of view, this requires measuring switching costs and quantifying the damages caused by all exploits in all possible configurations.
Secondly, they would also need to measure the adversary's cost of executing each exploit and their perceived gain for each successful attack.
As mentioned previously, this can be done using the CVSS cybersecurity metric and is investigated in \cite{sengupta2017game,sengupta2018moving}.
Finally, stochastic games assume that the player has knowledge of the current configuration.
This is justified as the adversary is assumed to be able to interact with it and the defender selected it.

In summary, we find:
\begin{itemize}
    \item \textbf{The Good.} Partial guidance on the ``where, when, and how'' question of MTD switching is provided, and the CVSS metric has been applied to leverage real world data in parameter calculation.
    \item \textbf{The Bad.}
    Solutions to game-theoretic models do not provide a switching rate, which would complete the specification of ``where, when, and how" of an MTD policy, and fail to formalize how the model should be recalibrated over time.
    \item \textbf{The Ugly.}
    MTD is typically an automated domain where switching decisions are made algorithmically. 
    This facilitates many interactions between attackers and a defender, where the defender has the opportunity to learn about the attackers' identity, intentions, and capabilities (i.e. through honeypots).
    Future IDS models should explicitly consider these learning dynamics over time, across many adversaries simultaneously.
    
\end{itemize}

\subsection{Efficacy and Practicality of CTI Sharing Models}
The problem of CTI sharing is unique, as it is a non-adversarial model.
This has significant impact on the information structure, for example it is reasonable to know about other agents as they are potential CTI collaborators.
Typically, solutions of this type of model are computed from an authorities perspective, rather than individual firms.
That is, an authority could use a game-theoretic CTI model to compute a likely joint response to their policy before implementing it.
Thus, when we evaluate practicality, it is still necessary that the central authority be able to measure all the information needed to compute an equilibrium.
We now identify four attributes that an ideal CTI sharing model would have:
\begin{enumerate}
    \item Identify and model realistic scenarios of CTI sharing such as a trusted central entity or distributed firms;
    \item model the trade-offs of risk, trust, competitive advantage, and security gain among competing firms;
    \item incorporate real mechanisms of CTI sharing such as the Structured Information Expression (STIX) framework (i.e. CTI is not a continuous infinitely divisible quantity);
    \item quantifying cybersecurity gain of obtaining CTI and 
loss of a breach.
\end{enumerate}

The problem of CTI sharing has been studied using Normal form, Differential games, Evolutionary games, and Coalitional games.
Using Normal-form games, the authors assume firms decide either to share or not.
Firms incur a mutual sharing benefit if they decide to share, as well as an infrastructure cost for effort required to package information.
In the Differential game setup \cite{gao2016differential} firms share CTI as a continuous quantity that impacts the derivative of their security.
Both the situation with 2 firms sharing and with a trusted third party have been considered in this context.
With respect to Evolutionary games, a similar setup is considered, but now firms choose to opt in and out of sharing \cite{tosh2015evolutionary} as a discrete choice and update their decision based on the average welfare of a firm deciding to share.
In Coalitional games, firms strategically decide to form coalitions based on a participation fee and security benefit from joining a coalition.
This addresses the free-rider problem prevalent in the CTI sharing domain.

We now evaluate these four approaches with the four modeling objectives that we have established.
First, we find that there is significant variety in the structure of the CTI sharing scenarios considered.
Normal form games and coalitional games have been applied to study the case of distributed firms, while evolutionary and Differential games have studied the situation with a trusted third party.
Additionally, it is worth noting that coalitional games not only model a situation with distributed firms but also capture endogenous coalition building.
Next, we also find that significant effort has been put into modeling the trade-off of competing firms, in the form of both normal form \cite{collins2021paying} and differential games \cite{gao2016differential}, as well as analysis of complementary firms in the case of differential games \cite{gao2014game}.
However, we find that no model achieves either of the remaining objectives.
Specifically, no model we reviewed considered CTI to be a real discrete object with practical considerations.
These considerations are: once threat data is obtained, it must be processed into a shareable format, possibly processed to redact sensitive information; and
the discovery process is likely stochastic and dynamic with respect to time, a dimension no model has explicitly considered.
We recommend that future models explicitly incorporate a model like STIX, which is a real world data structure for CTI, enhancing the efficacy of CTI models.
Additionally, to generate useful models the parameter data must be measured from the real world, so we recommend future models consider how to include assessments of likely cyber-breach damage and the value of receiving CTI.

We now consider the practicality of the solution concepts associated with differential games, normal form games, evolutionary games, and coalitional games.
From Table~\ref{tab:knowledge by model}, all four of these models assume knowledge of $A,N,U$.
Uniquely because CTI sharing is not an adversarial model, knowledge of the action set and player set $A,N$ is justified.
Intuitively, firms have knowledge of what other firms they might share with, and the action set encodes deciding how much CTI to share.
The utility function is less clear because firms must gain knowledge both of how to evaluate increased security by receiving CTI, the risks, and the infrastructure costs associated with sharing.
Critically, there are several orthogonal issues baked into the problem of finding a utility function:
(i) all works consider CTI to be an abstract and unitless quantity, and (ii) no work considers real mechanisms for how receiving CTI would increase security posture.
To remedy these issues, we first recommend that the models use real cybersecurity frameworks for CTI sharing, such as the STIX framework, which provides a standard for producing shareable data of cybersecurity events. 
Second, we recommend that future work use an approach inspired by CVSS where exploits and vulnerabilities can be rated qualitatively in order to produce quantitative results that can be reasoned with.
The above discussion resolves how a firm should construct its own utility function, but to calculate the appropriate equilibrium it must also learn the other firm's utility functions.
Because estimating how other firms will evaluate situations is inherently uncertain, future work should include analysis over a class of possible utility functions $\mathcal{U}$.

Additionally, the Differential game approaches assumes that an observable state exists in the model, typically representing an emergent security level based on the behavior of the firms.
This security level is calculated from the expected security outcomes given a simple attacker model \cite{gao2016differential}.
Although it is reasonable to assume that the security level will increase given increases in investment or CTI sharing, it is not clear what the structure of the function should be that governs this.
For example, investing in cybersecurity posture could be implemented in many ways (i.e. hiring more experts, modernizing infrastructure, bug bounty programs, etc.) all of which may impact the emergent security level differently.
Unfortunately, no Differential game model considers real mechanisms of how CTI exchanging hardens cybersecurity posture. 
Therefore, we propose that future works be more specific about what measures will be implemented with increased with cyber investment and model the mechanics of such a measure explicitly in the state function.
This would increase both the practicality of the model as there would be a real world basis to learn the parameters of the state function.

In summary, we find:
\begin{itemize}
    \item \textbf{The Good.} A wide variety of CTI sharing situations are modeled, and the economic considerations between firms have seen significant study.
    \item \textbf{The Bad.} Models fail to leverage real world data to evaluate the security gain and risk of CTI exchanges.
    \item \textbf{The Ugly.} Because the existing work assumes that CTI is an abstract and unit-less quantity, the usefulness of the results is greatly diminished.
    This is because the fundamental interaction being considered, the exchange of CTI, cannot be compared between modeled and real interactions.
    We recommend incorporating the STIX framework and other real considerations of CTI sharing (such as a stochastic discovery process following a cybersecurity event) to increase the efficacy and practicality of the model/solution.
\end{itemize}


\subsection{Other Considerations: Rationality and Collusion}
The Nash condition and its variants assume {\em rationality} of agents, which refers to the players' willingness and ability to optimize their utility functions (e.g., a player sometimes can get more utility than the equilibrium utility should the other agents fail to behave rationally). However, rationality can fail for many reasons (i.e. humans struggle to compute a utility function \cite{kahneman1982judgment}).
Thus, Nash equilibrium may not adequately model real world attacker and defender behaviors.

To resolve this deficit, future research should investigate models that can accommodate real world cybersecurity decision-making and rationality.
For this purpose, future studies should clearly state the rationality that is expected by a model (e.g., what kind of capability is needed for the defender to compute the utility when the attacker is not rational).
This may require specifying the tasks or levels of abstraction that are handled automatically, handled with algorithmic assistance, and handled completely \emph{ad hoc}. 

Additionally, the Nash condition has no intrinsic robustness to collusion.
This issue has received little attention because most studies focus on two-player games, but is important when attackers can collude 
for their mutual benefit.

To resolve this deficit, future research should consider multiple attackers.
There are some initial efforts to generalize the Nash equilibrium to provide robustness against coalitional deviations, such as the coalition-proof Nash equilibrium \cite{bernheim1987coalition} and the $k$-resilient Nash equilibrium \cite{abraham2006distributed}.
However, these studies are limited because they assume that collusion can be formed between all subsets of players, which may not apply to the cybersecurity domain.
We envision multiple kinds of collusion-resistant equilibrium definitions
because economically vs. politically-motivated attackers may exhibit different kinds of collusion between the attackers.

\section{Conclusion}
\label{sec:conclusion}

We have presented a framework to systematize game-theoretic cybersecurity studies. 
The framework also offers a systematic way to articulate game-theoretic cybersecurity models in future studies, including their assumptions. The framework would make it relatively easy for researchers and practitioners to assess the usefulness and validity of a model. 
Moreover, we assess the efficacy and practicality of existing game-theoretic modeling approaches for each of the four identified applications and provide recommendations in both areas.
We hope that the identified shortcomings and suggestions will inspire community endeavors to fulfill the potential of game theory in tackling cybersecurity decision-making problems. 

One limitation of our study is that we only consider four representative applications of game theory to cybersecurity. 
Although this does not undermine the systematization, it would be interesting to apply the framework to analyze other applications of game theory to cybersecurity. 

\bibliographystyle{ACM-Reference-Format}
\bibliography{references,metrics,Philiplibrary}


\begin{thebibliography}{110}


\ifx \showCODEN    \undefined \def \showCODEN     #1{\unskip}     \fi
\ifx \showDOI      \undefined \def \showDOI       #1{#1}\fi
\ifx \showISBNx    \undefined \def \showISBNx     #1{\unskip}     \fi
\ifx \showISBNxiii \undefined \def \showISBNxiii  #1{\unskip}     \fi
\ifx \showISSN     \undefined \def \showISSN      #1{\unskip}     \fi
\ifx \showLCCN     \undefined \def \showLCCN      #1{\unskip}     \fi
\ifx \shownote     \undefined \def \shownote      #1{#1}          \fi
\ifx \showarticletitle \undefined \def \showarticletitle #1{#1}   \fi
\ifx \showURL      \undefined \def \showURL       {\relax}        \fi
\providecommand\bibfield[2]{#2}
\providecommand\bibinfo[2]{#2}
\providecommand\natexlab[1]{#1}
\providecommand\showeprint[2][]{arXiv:#2}

\bibitem[Abass et~al\mbox{.}(2017)]%
        {abass2017evolutionary}
\bibfield{author}{\bibinfo{person}{Ahmed A~Alabdel Abass}, \bibinfo{person}{Liang Xiao}, \bibinfo{person}{Narayan~B Mandayam}, {and} \bibinfo{person}{Zoran Gajic}.} \bibinfo{year}{2017}\natexlab{}.
\newblock \showarticletitle{Evolutionary game theoretic analysis of advanced persistent threats against cloud storage}.
\newblock \bibinfo{journal}{\emph{IEEE Access}}  \bibinfo{volume}{5} (\bibinfo{year}{2017}), \bibinfo{pages}{8482--8491}.
\newblock


\bibitem[Abraham et~al\mbox{.}(2006)]%
        {abraham2006distributed}
\bibfield{author}{\bibinfo{person}{Ittai Abraham}, \bibinfo{person}{Danny Dolev}, \bibinfo{person}{Rica Gonen}, {and} \bibinfo{person}{Joe Halpern}.} \bibinfo{year}{2006}\natexlab{}.
\newblock \showarticletitle{Distributed computing meets game theory: robust mechanisms for rational secret sharing and multiparty computation}. In \bibinfo{booktitle}{\emph{Proceedings of the twenty-fifth annual ACM symposium on Principles of distributed computing}}. \bibinfo{pages}{53--62}.
\newblock


\bibitem[Abusitta et~al\mbox{.}(2018)]%
        {abusitta2018trust}
\bibfield{author}{\bibinfo{person}{Adel Abusitta}, \bibinfo{person}{Martine Bellaiche}, {and} \bibinfo{person}{Michel Dagenais}.} \bibinfo{year}{2018}\natexlab{}.
\newblock \showarticletitle{A trust-based game theoretical model for cooperative intrusion detection in multi-cloud environments}. In \bibinfo{booktitle}{\emph{2018 21st Conference on Innovation in Clouds, Internet and Networks and Workshops (ICIN)}}. IEEE, \bibinfo{pages}{1--8}.
\newblock


\bibitem[Adami and Hintze(2013)]%
        {adami2013evolutionary}
\bibfield{author}{\bibinfo{person}{Christoph Adami} {and} \bibinfo{person}{Arend Hintze}.} \bibinfo{year}{2013}\natexlab{}.
\newblock \showarticletitle{Evolutionary instability of zero-determinant strategies demonstrates that winning is not everything}.
\newblock \bibinfo{journal}{\emph{Nature communications}} \bibinfo{volume}{4}, \bibinfo{number}{1} (\bibinfo{year}{2013}), \bibinfo{pages}{2193}.
\newblock


\bibitem[Agah et~al\mbox{.}(2004)]%
        {agah2004intrusion}
\bibfield{author}{\bibinfo{person}{Afrand Agah}, \bibinfo{person}{Sajal~K Das}, \bibinfo{person}{Kalyan Basu}, {and} \bibinfo{person}{Mehran Asadi}.} \bibinfo{year}{2004}\natexlab{}.
\newblock \showarticletitle{Intrusion detection in sensor networks: A non-cooperative game approach}. In \bibinfo{booktitle}{\emph{Third IEEE International Symposium on Network Computing and Applications, 2004.(NCA 2004). Proceedings.}} IEEE, \bibinfo{pages}{343--346}.
\newblock


\bibitem[Alpcan and Basar(2003)]%
        {alpcan2003game}
\bibfield{author}{\bibinfo{person}{Tansu Alpcan} {and} \bibinfo{person}{Tamer Basar}.} \bibinfo{year}{2003}\natexlab{}.
\newblock \showarticletitle{A game theoretic approach to decision and analysis in network intrusion detection}. In \bibinfo{booktitle}{\emph{42nd IEEE International Conference on Decision and Control (IEEE Cat. No. 03CH37475)}}, Vol.~\bibinfo{volume}{3}. IEEE, \bibinfo{pages}{2595--2600}.
\newblock


\bibitem[Alpcan and Basar(2004)]%
        {alpcan2004game}
\bibfield{author}{\bibinfo{person}{Tansu Alpcan} {and} \bibinfo{person}{Tamer Basar}.} \bibinfo{year}{2004}\natexlab{}.
\newblock \showarticletitle{A game theoretic analysis of intrusion detection in access control systems}. In \bibinfo{booktitle}{\emph{2004 43rd IEEE Conference on Decision and Control (CDC)(IEEE Cat. No. 04CH37601)}}, Vol.~\bibinfo{volume}{2}. IEEE, \bibinfo{pages}{1568--1573}.
\newblock


\bibitem[Alpcan and Basar(2006)]%
        {alpcan2006intrusion}
\bibfield{author}{\bibinfo{person}{Tansu Alpcan} {and} \bibinfo{person}{Tamer Basar}.} \bibinfo{year}{2006}\natexlab{}.
\newblock \showarticletitle{An intrusion detection game with limited observations}. In \bibinfo{booktitle}{\emph{12th Int. Symp. on Dynamic Games and Applications, Sophia Antipolis, France}}, Vol.~\bibinfo{volume}{26}.
\newblock


\bibitem[Anwar et~al\mbox{.}(2017)]%
        {anwar2017dynamic}
\bibfield{author}{\bibinfo{person}{Ahmed~H Anwar}, \bibinfo{person}{George Atia}, {and} \bibinfo{person}{Mina Guirguis}.} \bibinfo{year}{2017}\natexlab{}.
\newblock \showarticletitle{Dynamic game-theoretic defense approach against stealthy jamming attacks in wireless networks}. In \bibinfo{booktitle}{\emph{2017 55th Annual Allerton Conference on Communication, Control, and Computing (Allerton)}}. IEEE, \bibinfo{pages}{252--258}.
\newblock


\bibitem[Anwar et~al\mbox{.}(2020)]%
        {anwar2020honeypot}
\bibfield{author}{\bibinfo{person}{Ahmed~H Anwar}, \bibinfo{person}{Charles Kamhoua}, {and} \bibinfo{person}{Nandi Leslie}.} \bibinfo{year}{2020}\natexlab{}.
\newblock \showarticletitle{Honeypot allocation over attack graphs in cyber deception games}. In \bibinfo{booktitle}{\emph{2020 International Conference on Computing, Networking and Communications (ICNC)}}. IEEE, \bibinfo{pages}{502--506}.
\newblock


\bibitem[Anwar and Kamhoua(2022)]%
        {anwar2022cyber}
\bibfield{author}{\bibinfo{person}{Ahmed~H Anwar} {and} \bibinfo{person}{Charles~A Kamhoua}.} \bibinfo{year}{2022}\natexlab{}.
\newblock \showarticletitle{Cyber Deception using Honeypot Allocation and Diversity: A Game Theoretic Approach}. In \bibinfo{booktitle}{\emph{2022 IEEE 19th Annual Consumer Communications \& Networking Conference (CCNC)}}. IEEE, \bibinfo{pages}{543--549}.
\newblock


\bibitem[Anwar et~al\mbox{.}(2022)]%
        {anwar2022honeypot}
\bibfield{author}{\bibinfo{person}{Ahmed~H Anwar}, \bibinfo{person}{Charles~A Kamhoua}, \bibinfo{person}{Nandi~O Leslie}, {and} \bibinfo{person}{Christopher Kiekintveld}.} \bibinfo{year}{2022}\natexlab{}.
\newblock \showarticletitle{Honeypot Allocation for Cyber Deception Under Uncertainty}.
\newblock \bibinfo{journal}{\emph{IEEE Transactions on Network and Service Management}} \bibinfo{volume}{19}, \bibinfo{number}{3} (\bibinfo{year}{2022}), \bibinfo{pages}{3438--3452}.
\newblock


\bibitem[Aydeger et~al\mbox{.}(2021)]%
        {aydeger2021strategic}
\bibfield{author}{\bibinfo{person}{Abdullah Aydeger}, \bibinfo{person}{Mohammad~Hossein Manshaei}, \bibinfo{person}{Mohammad~Ashiqur Rahman}, {and} \bibinfo{person}{Kemal Akkaya}.} \bibinfo{year}{2021}\natexlab{}.
\newblock \showarticletitle{Strategic defense against stealthy link flooding attacks: A signaling game approach}.
\newblock \bibinfo{journal}{\emph{IEEE Transactions on Network Science and Engineering}} \bibinfo{volume}{8}, \bibinfo{number}{1} (\bibinfo{year}{2021}), \bibinfo{pages}{751--764}.
\newblock


\bibitem[Basak et~al\mbox{.}(2019)]%
        {basak2019identifying}
\bibfield{author}{\bibinfo{person}{Anjon Basak}, \bibinfo{person}{Charles Kamhoua}, \bibinfo{person}{Sridhar Venkatesan}, \bibinfo{person}{Marcus Gutierrez}, \bibinfo{person}{Ahmed~H Anwar}, {and} \bibinfo{person}{Christopher Kiekintveld}.} \bibinfo{year}{2019}\natexlab{}.
\newblock \showarticletitle{Identifying stealthy attackers in a game theoretic framework using deception}. In \bibinfo{booktitle}{\emph{Decision and Game Theory for Security: 10th International Conference, GameSec 2019, Stockholm, Sweden, October 30--November 1, 2019, Proceedings 10}}. Springer, \bibinfo{pages}{21--32}.
\newblock


\bibitem[Bernheim et~al\mbox{.}(1987)]%
        {bernheim1987coalition}
\bibfield{author}{\bibinfo{person}{B~Douglas Bernheim}, \bibinfo{person}{Bezalel Peleg}, {and} \bibinfo{person}{Michael~D Whinston}.} \bibinfo{year}{1987}\natexlab{}.
\newblock \showarticletitle{Coalition-proof nash equilibria i. concepts}.
\newblock \bibinfo{journal}{\emph{Journal of economic theory}} \bibinfo{volume}{42}, \bibinfo{number}{1} (\bibinfo{year}{1987}), \bibinfo{pages}{1--12}.
\newblock


\bibitem[Bertsekas(2012)]%
        {bertsekas2012dynamic}
\bibfield{author}{\bibinfo{person}{Dimitri Bertsekas}.} \bibinfo{year}{2012}\natexlab{}.
\newblock \bibinfo{booktitle}{\emph{Dynamic programming and optimal control: Volume I}}. Vol.~\bibinfo{volume}{4}.
\newblock \bibinfo{publisher}{Athena scientific}.
\newblock


\bibitem[Bouhaddi et~al\mbox{.}(2018)]%
        {bouhaddi2018efficient}
\bibfield{author}{\bibinfo{person}{Myria Bouhaddi}, \bibinfo{person}{Mohammed~Said Radjef}, {and} \bibinfo{person}{Kamel Adi}.} \bibinfo{year}{2018}\natexlab{}.
\newblock \showarticletitle{An efficient intrusion detection in resource-constrained mobile ad-hoc networks}.
\newblock \bibinfo{journal}{\emph{Computers \& Security}}  \bibinfo{volume}{76} (\bibinfo{year}{2018}), \bibinfo{pages}{156--177}.
\newblock


\bibitem[Boumkheld et~al\mbox{.}(2019)]%
        {boumkheld2019honeypot}
\bibfield{author}{\bibinfo{person}{Nadia Boumkheld}, \bibinfo{person}{Sakshyam Panda}, \bibinfo{person}{Stefan Rass}, {and} \bibinfo{person}{Emmanouil Panaousis}.} \bibinfo{year}{2019}\natexlab{}.
\newblock \showarticletitle{Honeypot type selection games for smart grid networks}. In \bibinfo{booktitle}{\emph{Decision and Game Theory for Security: 10th International Conference, GameSec 2019, Stockholm, Sweden, October 30--November 1, 2019, Proceedings 10}}. Springer, \bibinfo{pages}{85--96}.
\newblock


\bibitem[Boyd and Vandenberghe(2004)]%
        {boyd2004convex}
\bibfield{author}{\bibinfo{person}{Stephen~P Boyd} {and} \bibinfo{person}{Lieven Vandenberghe}.} \bibinfo{year}{2004}\natexlab{}.
\newblock \bibinfo{booktitle}{\emph{Convex optimization}}.
\newblock \bibinfo{publisher}{Cambridge university press}.
\newblock


\bibitem[Carter et~al\mbox{.}(2014)]%
        {carter2014game}
\bibfield{author}{\bibinfo{person}{Kevin~M Carter}, \bibinfo{person}{James~F Riordan}, {and} \bibinfo{person}{Hamed Okhravi}.} \bibinfo{year}{2014}\natexlab{}.
\newblock \showarticletitle{A game theoretic approach to strategy determination for dynamic platform defenses}. In \bibinfo{booktitle}{\emph{Proceedings of the first ACM workshop on moving target defense}}. \bibinfo{pages}{21--30}.
\newblock


\bibitem[Chen and Leneutre(2009)]%
        {chen2009game}
\bibfield{author}{\bibinfo{person}{Lin Chen} {and} \bibinfo{person}{Jean Leneutre}.} \bibinfo{year}{2009}\natexlab{}.
\newblock \showarticletitle{A game theoretical framework on intrusion detection in heterogeneous networks}.
\newblock \bibinfo{journal}{\emph{IEEE Transactions on Information Forensics and Security}} \bibinfo{volume}{4}, \bibinfo{number}{2} (\bibinfo{year}{2009}), \bibinfo{pages}{165--178}.
\newblock


\bibitem[Cho et~al\mbox{.}(2020)]%
        {cho2020toward}
\bibfield{author}{\bibinfo{person}{Jin-Hee Cho}, \bibinfo{person}{Dilli~P Sharma}, \bibinfo{person}{Hooman Alavizadeh}, \bibinfo{person}{Seunghyun Yoon}, \bibinfo{person}{Noam Ben-Asher}, \bibinfo{person}{Terrence~J Moore}, \bibinfo{person}{Dong~Seong Kim}, \bibinfo{person}{Hyuk Lim}, {and} \bibinfo{person}{Frederica~F Nelson}.} \bibinfo{year}{2020}\natexlab{}.
\newblock \showarticletitle{Toward proactive, adaptive defense: A survey on moving target defense}.
\newblock \bibinfo{journal}{\emph{IEEE Communications Surveys \& Tutorials}} \bibinfo{volume}{22}, \bibinfo{number}{1} (\bibinfo{year}{2020}), \bibinfo{pages}{709--745}.
\newblock


\bibitem[Chowdhary and Sailik~Sengupta(2019)]%
        {chowdhary2018markov}
\bibfield{author}{\bibinfo{person}{Ankur Chowdhary} {and} \bibinfo{person}{Subbarao~Kambhampati Sailik~Sengupta, Dijiang~Huang}.} \bibinfo{year}{2019}\natexlab{}.
\newblock \showarticletitle{Markov game modeling of moving target defense for strategic detection of threats in cloud networks}.
\newblock \bibinfo{journal}{\emph{Proc. AAAI Workshop Artificial Intelligence Cyber Security (AICS)}} (\bibinfo{year}{2019}).
\newblock


\bibitem[Collins et~al\mbox{.}(2021)]%
        {collins2021paying}
\bibfield{author}{\bibinfo{person}{Brandon Collins}, \bibinfo{person}{Shouhuai Xu}, {and} \bibinfo{person}{Philip~N Brown}.} \bibinfo{year}{2021}\natexlab{}.
\newblock \showarticletitle{Paying Firms to Share Cyber Threat Intelligence}. In \bibinfo{booktitle}{\emph{International Conference on Decision and Game Theory for Security}}. Springer, \bibinfo{pages}{365--377}.
\newblock


\bibitem[Conitzer and Sandholm(2006)]%
        {conitzer2006computing}
\bibfield{author}{\bibinfo{person}{Vincent Conitzer} {and} \bibinfo{person}{Tuomas Sandholm}.} \bibinfo{year}{2006}\natexlab{}.
\newblock \showarticletitle{Computing the optimal strategy to commit to}. In \bibinfo{booktitle}{\emph{Proceedings of the 7th ACM conference on Electronic commerce}}. \bibinfo{pages}{82--90}.
\newblock


\bibitem[Diamantoulakis et~al\mbox{.}(2020)]%
        {diamantoulakis2020game}
\bibfield{author}{\bibinfo{person}{Panagiotis Diamantoulakis}, \bibinfo{person}{Christos Dalamagkas}, \bibinfo{person}{Panagiotis Radoglou-Grammatikis}, \bibinfo{person}{Panagiotis Sarigiannidis}, {and} \bibinfo{person}{George Karagiannidis}.} \bibinfo{year}{2020}\natexlab{}.
\newblock \showarticletitle{Game theoretic honeypot deployment in smart grid}.
\newblock \bibinfo{journal}{\emph{Sensors}} \bibinfo{volume}{20}, \bibinfo{number}{15} (\bibinfo{year}{2020}), \bibinfo{pages}{4199}.
\newblock


\bibitem[Do et~al\mbox{.}(2017)]%
        {do2017game}
\bibfield{author}{\bibinfo{person}{Cuong~T Do}, \bibinfo{person}{Nguyen~H Tran}, \bibinfo{person}{Choongseon Hong}, \bibinfo{person}{Charles~A Kamhoua}, \bibinfo{person}{Kevin~A Kwiat}, \bibinfo{person}{Erik Blasch}, \bibinfo{person}{Shaolei Ren}, \bibinfo{person}{Niki Pissinou}, {and} \bibinfo{person}{Sundaraja~Sitharama Iyengar}.} \bibinfo{year}{2017}\natexlab{}.
\newblock \showarticletitle{Game theory for cyber security and privacy}.
\newblock \bibinfo{journal}{\emph{ACM Computing Surveys (CSUR)}} \bibinfo{volume}{50}, \bibinfo{number}{2} (\bibinfo{year}{2017}), \bibinfo{pages}{1--37}.
\newblock


\bibitem[Du and Wang(2019)]%
        {du2019sdn}
\bibfield{author}{\bibinfo{person}{Miao Du} {and} \bibinfo{person}{Kun Wang}.} \bibinfo{year}{2019}\natexlab{}.
\newblock \showarticletitle{An SDN-enabled pseudo-honeypot strategy for distributed denial of service attacks in industrial Internet of Things}.
\newblock \bibinfo{journal}{\emph{IEEE Transactions on Industrial Informatics}} \bibinfo{volume}{16}, \bibinfo{number}{1} (\bibinfo{year}{2019}), \bibinfo{pages}{648--657}.
\newblock


\bibitem[Eaves(1971)]%
        {eaves1971linear}
\bibfield{author}{\bibinfo{person}{B~Curtis Eaves}.} \bibinfo{year}{1971}\natexlab{}.
\newblock \showarticletitle{The linear complementarity problem}.
\newblock \bibinfo{journal}{\emph{Management science}} \bibinfo{volume}{17}, \bibinfo{number}{9} (\bibinfo{year}{1971}), \bibinfo{pages}{612--634}.
\newblock


\bibitem[Farhang and Grossklags(2016)]%
        {farhang2016flipleakage}
\bibfield{author}{\bibinfo{person}{Sadegh Farhang} {and} \bibinfo{person}{Jens Grossklags}.} \bibinfo{year}{2016}\natexlab{}.
\newblock \showarticletitle{FlipLeakage: a game-theoretic approach to protect against stealthy attackers in the presence of information leakage}. In \bibinfo{booktitle}{\emph{International Conference on Decision and Game Theory for Security}}. Springer, \bibinfo{pages}{195--214}.
\newblock


\bibitem[Feng et~al\mbox{.}(2017a)]%
        {feng2017signaling}
\bibfield{author}{\bibinfo{person}{Xiaotao Feng}, \bibinfo{person}{Zizhan Zheng}, \bibinfo{person}{Derya Cansever}, \bibinfo{person}{Ananthram Swami}, {and} \bibinfo{person}{Prasant Mohapatra}.} \bibinfo{year}{2017}\natexlab{a}.
\newblock \showarticletitle{A signaling game model for moving target defense}. In \bibinfo{booktitle}{\emph{IEEE INFOCOM 2017-IEEE conference on computer communications}}. IEEE, \bibinfo{pages}{1--9}.
\newblock


\bibitem[Feng et~al\mbox{.}(2015)]%
        {feng2015stealthy}
\bibfield{author}{\bibinfo{person}{Xiaotao Feng}, \bibinfo{person}{Zizhan Zheng}, \bibinfo{person}{Pengfei Hu}, \bibinfo{person}{Derya Cansever}, {and} \bibinfo{person}{Prasant Mohapatra}.} \bibinfo{year}{2015}\natexlab{}.
\newblock \showarticletitle{Stealthy attacks meets insider threats: A three-player game model}. In \bibinfo{booktitle}{\emph{MILCOM 2015-2015 IEEE Military Communications Conference}}. IEEE, \bibinfo{pages}{25--30}.
\newblock


\bibitem[Feng et~al\mbox{.}(2017b)]%
        {feng2017stackelberg}
\bibfield{author}{\bibinfo{person}{Xiaotao Feng}, \bibinfo{person}{Zizhan Zheng}, \bibinfo{person}{Prasant Mohapatra}, {and} \bibinfo{person}{Derya Cansever}.} \bibinfo{year}{2017}\natexlab{b}.
\newblock \showarticletitle{A stackelberg game and markov modeling of moving target defense}. In \bibinfo{booktitle}{\emph{Decision and Game Theory for Security: 8th International Conference, GameSec 2017, Vienna, Austria, October 23-25, 2017, Proceedings}}. Springer, \bibinfo{pages}{315--335}.
\newblock


\bibitem[Fotiadis et~al\mbox{.}(2020)]%
        {fotiadis2020constrained}
\bibfield{author}{\bibinfo{person}{Filippos Fotiadis}, \bibinfo{person}{Aris Kanellopoulos}, {and} \bibinfo{person}{Kyriakos~G Vamvoudakis}.} \bibinfo{year}{2020}\natexlab{}.
\newblock \showarticletitle{Constrained differential games for secure decision-making against stealthy attacks}. In \bibinfo{booktitle}{\emph{2020 American Control Conference (ACC)}}. IEEE, \bibinfo{pages}{4658--4663}.
\newblock


\bibitem[Gao and Zhong(2016)]%
        {gao2016differential}
\bibfield{author}{\bibinfo{person}{Xing Gao} {and} \bibinfo{person}{Weijun Zhong}.} \bibinfo{year}{2016}\natexlab{}.
\newblock \showarticletitle{A differential game approach to security investment and information sharing in a competitive environment}.
\newblock \bibinfo{journal}{\emph{IIE Transactions}} \bibinfo{volume}{48}, \bibinfo{number}{6} (\bibinfo{year}{2016}), \bibinfo{pages}{511--526}.
\newblock


\bibitem[Gao et~al\mbox{.}(2014)]%
        {gao2014game}
\bibfield{author}{\bibinfo{person}{Xing Gao}, \bibinfo{person}{Weijun Zhong}, {and} \bibinfo{person}{Shue Mei}.} \bibinfo{year}{2014}\natexlab{}.
\newblock \showarticletitle{A game-theoretic analysis of information sharing and security investment for complementary firms}.
\newblock \bibinfo{journal}{\emph{Journal of the Operational Research Society}} \bibinfo{volume}{65}, \bibinfo{number}{11} (\bibinfo{year}{2014}), \bibinfo{pages}{1682--1691}.
\newblock


\bibitem[Gill et~al\mbox{.}(2020)]%
        {gill2020gtm}
\bibfield{author}{\bibinfo{person}{Komal~Singh Gill}, \bibinfo{person}{Sharad Saxena}, {and} \bibinfo{person}{Anju Sharma}.} \bibinfo{year}{2020}\natexlab{}.
\newblock \showarticletitle{GTM-CSec: game theoretic model for cloud security based on IDS and honeypot}.
\newblock \bibinfo{journal}{\emph{Computers \& Security}}  \bibinfo{volume}{92} (\bibinfo{year}{2020}), \bibinfo{pages}{101732}.
\newblock


\bibitem[Goldreich(2001)]%
        {GoldreichBookVol1}
\bibfield{author}{\bibinfo{person}{O. Goldreich}.} \bibinfo{year}{2001}\natexlab{}.
\newblock \bibinfo{booktitle}{\emph{The Foundations of Cryptography}}. Vol.~\bibinfo{volume}{1}.
\newblock \bibinfo{publisher}{Cambridge University Press}.
\newblock


\bibitem[Gothawal and Nagaraj(2020)]%
        {gothawal2020anomaly}
\bibfield{author}{\bibinfo{person}{Deepali~Bankatsingh Gothawal} {and} \bibinfo{person}{SV Nagaraj}.} \bibinfo{year}{2020}\natexlab{}.
\newblock \showarticletitle{Anomaly-based intrusion detection system in RPL by applying stochastic and evolutionary game models over IoT environment}.
\newblock \bibinfo{journal}{\emph{Wireless Personal Communications}}  \bibinfo{volume}{110} (\bibinfo{year}{2020}), \bibinfo{pages}{1323--1344}.
\newblock


\bibitem[Guo et~al\mbox{.}(2018)]%
        {guo2018incentive}
\bibfield{author}{\bibinfo{person}{Yunchuan Guo}, \bibinfo{person}{Han Zhang}, \bibinfo{person}{Lingcui Zhang}, \bibinfo{person}{Liang Fang}, {and} \bibinfo{person}{Fenghua Li}.} \bibinfo{year}{2018}\natexlab{}.
\newblock \showarticletitle{Incentive mechanism for cooperative intrusion detection: an evolutionary game approach}. In \bibinfo{booktitle}{\emph{Computational Science--ICCS 2018: 18th International Conference, Wuxi, China, June 11--13, 2018, Proceedings, Part I 18}}. Springer, \bibinfo{pages}{83--97}.
\newblock


\bibitem[Han et~al\mbox{.}(2019)]%
        {han2019intrusion}
\bibfield{author}{\bibinfo{person}{Lansheng Han}, \bibinfo{person}{Man Zhou}, \bibinfo{person}{Wenjing Jia}, \bibinfo{person}{Zakaria Dalil}, {and} \bibinfo{person}{Xingbo Xu}.} \bibinfo{year}{2019}\natexlab{}.
\newblock \showarticletitle{Intrusion detection model of wireless sensor networks based on game theory and an autoregressive model}.
\newblock \bibinfo{journal}{\emph{Information sciences}}  \bibinfo{volume}{476} (\bibinfo{year}{2019}), \bibinfo{pages}{491--504}.
\newblock


\bibitem[Hart and Mansour(2010)]%
        {Hart2010}
\bibfield{author}{\bibinfo{person}{Sergiu Hart} {and} \bibinfo{person}{Yishay Mansour}.} \bibinfo{year}{2010}\natexlab{}.
\newblock \showarticletitle{{How long to equilibrium? The communication complexity of uncoupled equilibrium procedures}}.
\newblock \bibinfo{journal}{\emph{Games and Economic Behavior}} \bibinfo{volume}{69}, \bibinfo{number}{1} (\bibinfo{year}{2010}), \bibinfo{pages}{107--126}.
\newblock
\showISSN{08998256}
\urldef\tempurl%
\url{https://doi.org/10.1016/j.geb.2007.12.002}
\showDOI{\tempurl}


\bibitem[Hausken(2007)]%
        {hausken2007information}
\bibfield{author}{\bibinfo{person}{Kjell Hausken}.} \bibinfo{year}{2007}\natexlab{}.
\newblock \showarticletitle{Information sharing among firms and cyber attacks}.
\newblock \bibinfo{journal}{\emph{Journal of Accounting and Public Policy}} \bibinfo{volume}{26}, \bibinfo{number}{6} (\bibinfo{year}{2007}), \bibinfo{pages}{639--688}.
\newblock


\bibitem[Hor{\'a}k et~al\mbox{.}(2019)]%
        {horak2019optimizing}
\bibfield{author}{\bibinfo{person}{Karel Hor{\'a}k}, \bibinfo{person}{Branislav Bo{\v{s}}ansk{\`y}}, \bibinfo{person}{Petr Tom{\'a}{\v{s}}ek}, \bibinfo{person}{Christopher Kiekintveld}, {and} \bibinfo{person}{Charles Kamhoua}.} \bibinfo{year}{2019}\natexlab{}.
\newblock \showarticletitle{Optimizing honeypot strategies against dynamic lateral movement using partially observable stochastic games}.
\newblock \bibinfo{journal}{\emph{Computers \& Security}}  \bibinfo{volume}{87} (\bibinfo{year}{2019}), \bibinfo{pages}{101579}.
\newblock


\bibitem[Hu et~al\mbox{.}(2015)]%
        {hu2015dynamic}
\bibfield{author}{\bibinfo{person}{Pengfei Hu}, \bibinfo{person}{Hongxing Li}, \bibinfo{person}{Hao Fu}, \bibinfo{person}{Derya Cansever}, {and} \bibinfo{person}{Prasant Mohapatra}.} \bibinfo{year}{2015}\natexlab{}.
\newblock \showarticletitle{Dynamic defense strategy against advanced persistent threat with insiders}. In \bibinfo{booktitle}{\emph{2015 IEEE Conference on Computer Communications (INFOCOM)}}. IEEE, \bibinfo{pages}{747--755}.
\newblock


\bibitem[Hu et~al\mbox{.}(2017)]%
        {hu2017defense}
\bibfield{author}{\bibinfo{person}{Qing Hu}, \bibinfo{person}{Shichao Lv}, \bibinfo{person}{Zhiqiang Shi}, \bibinfo{person}{Limin Sun}, {and} \bibinfo{person}{Liang Xiao}.} \bibinfo{year}{2017}\natexlab{}.
\newblock \showarticletitle{Defense against advanced persistent threats with expert system for internet of things}. In \bibinfo{booktitle}{\emph{Wireless Algorithms, Systems, and Applications: 12th International Conference, WASA 2017, Guilin, China, June 19-21, 2017, Proceedings 12}}. Springer, \bibinfo{pages}{326--337}.
\newblock


\bibitem[Huang and Zhu(2018)]%
        {huang2018analysis}
\bibfield{author}{\bibinfo{person}{Linan Huang} {and} \bibinfo{person}{Quanyan Zhu}.} \bibinfo{year}{2018}\natexlab{}.
\newblock \showarticletitle{Analysis and computation of adaptive defense strategies against advanced persistent threats for cyber-physical systems}. In \bibinfo{booktitle}{\emph{Decision and Game Theory for Security: 9th International Conference, GameSec 2018, Seattle, WA, USA, October 29--31, 2018, Proceedings 9}}. Springer, \bibinfo{pages}{205--226}.
\newblock


\bibitem[Huang and Zhu(2019)]%
        {huang2019adaptive}
\bibfield{author}{\bibinfo{person}{Linan Huang} {and} \bibinfo{person}{Quanyan Zhu}.} \bibinfo{year}{2019}\natexlab{}.
\newblock \showarticletitle{Adaptive strategic cyber defense for advanced persistent threats in critical infrastructure networks}.
\newblock \bibinfo{journal}{\emph{ACM SIGMETRICS Performance Evaluation Review}} \bibinfo{volume}{46}, \bibinfo{number}{2} (\bibinfo{year}{2019}), \bibinfo{pages}{52--56}.
\newblock


\bibitem[Huang and Zhu(2020)]%
        {huang2020dynamic}
\bibfield{author}{\bibinfo{person}{Linan Huang} {and} \bibinfo{person}{Quanyan Zhu}.} \bibinfo{year}{2020}\natexlab{}.
\newblock \showarticletitle{A dynamic games approach to proactive defense strategies against advanced persistent threats in cyber-physical systems}.
\newblock \bibinfo{journal}{\emph{Computers \& Security}}  \bibinfo{volume}{89} (\bibinfo{year}{2020}), \bibinfo{pages}{101660}.
\newblock


\bibitem[Jajodia et~al\mbox{.}(2018)]%
        {jajodia2018share}
\bibfield{author}{\bibinfo{person}{Sushil Jajodia}, \bibinfo{person}{Noseong Park}, \bibinfo{person}{Edoardo Serra}, {and} \bibinfo{person}{VS Subrahmanian}.} \bibinfo{year}{2018}\natexlab{}.
\newblock \showarticletitle{Share: A stackelberg honey-based adversarial reasoning engine}.
\newblock \bibinfo{journal}{\emph{ACM Transactions on Internet Technology (TOIT)}} \bibinfo{volume}{18}, \bibinfo{number}{3} (\bibinfo{year}{2018}), \bibinfo{pages}{1--41}.
\newblock


\bibitem[Jiang and Bai(2019)]%
        {jiang2019evaluation}
\bibfield{author}{\bibinfo{person}{John~Xuefeng Jiang} {and} \bibinfo{person}{Ge Bai}.} \bibinfo{year}{2019}\natexlab{}.
\newblock \showarticletitle{Evaluation of causes of protected health information breaches}.
\newblock \bibinfo{journal}{\emph{JAMA internal medicine}} \bibinfo{volume}{179}, \bibinfo{number}{2} (\bibinfo{year}{2019}), \bibinfo{pages}{265--267}.
\newblock


\bibitem[Kahneman et~al\mbox{.}(1982)]%
        {kahneman1982judgment}
\bibfield{author}{\bibinfo{person}{Daniel Kahneman}, \bibinfo{person}{Stewart~Paul Slovic}, \bibinfo{person}{Paul Slovic}, {and} \bibinfo{person}{Amos Tversky}.} \bibinfo{year}{1982}\natexlab{}.
\newblock \bibinfo{booktitle}{\emph{Judgment under uncertainty: Heuristics and biases}}.
\newblock \bibinfo{publisher}{Cambridge university press}.
\newblock


\bibitem[Kamhoua et~al\mbox{.}(2015)]%
        {kamhoua2015cyber}
\bibfield{author}{\bibinfo{person}{Charles Kamhoua}, \bibinfo{person}{Andrew Martin}, \bibinfo{person}{Deepak~K Tosh}, \bibinfo{person}{Kevin~A Kwiat}, \bibinfo{person}{Chad Heitzenrater}, {and} \bibinfo{person}{Shamik Sengupta}.} \bibinfo{year}{2015}\natexlab{}.
\newblock \showarticletitle{Cyber-threats information sharing in cloud computing: A game theoretic approach}. In \bibinfo{booktitle}{\emph{2015 IEEE 2nd International Conference on Cyber Security and Cloud Computing}}. IEEE, \bibinfo{pages}{382--389}.
\newblock


\bibitem[Karlin and Peres(2017)]%
        {karlin2017game}
\bibfield{author}{\bibinfo{person}{Anna~R Karlin} {and} \bibinfo{person}{Yuval Peres}.} \bibinfo{year}{2017}\natexlab{}.
\newblock \bibinfo{booktitle}{\emph{Game theory, alive}}. Vol.~\bibinfo{volume}{101}.
\newblock \bibinfo{publisher}{American Mathematical Soc.}
\newblock


\bibitem[Kiennert et~al\mbox{.}(2018)]%
        {kiennert2018survey}
\bibfield{author}{\bibinfo{person}{Christophe Kiennert}, \bibinfo{person}{Ziad Ismail}, \bibinfo{person}{Herve Debar}, {and} \bibinfo{person}{Jean Leneutre}.} \bibinfo{year}{2018}\natexlab{}.
\newblock \showarticletitle{A survey on game-theoretic approaches for intrusion detection and response optimization}.
\newblock \bibinfo{journal}{\emph{ACM Computing Surveys (CSUR)}} \bibinfo{volume}{51}, \bibinfo{number}{5} (\bibinfo{year}{2018}), \bibinfo{pages}{1--31}.
\newblock


\bibitem[La et~al\mbox{.}(2016)]%
        {la2016deceptive}
\bibfield{author}{\bibinfo{person}{Quang~Duy La}, \bibinfo{person}{Tony~QS Quek}, \bibinfo{person}{Jemin Lee}, \bibinfo{person}{Shi Jin}, {and} \bibinfo{person}{Hongbo Zhu}.} \bibinfo{year}{2016}\natexlab{}.
\newblock \showarticletitle{Deceptive attack and defense game in honeypot-enabled networks for the internet of things}.
\newblock \bibinfo{journal}{\emph{IEEE Internet of Things Journal}} \bibinfo{volume}{3}, \bibinfo{number}{6} (\bibinfo{year}{2016}), \bibinfo{pages}{1025--1035}.
\newblock


\bibitem[Laszka et~al\mbox{.}(2014)]%
        {laszka2014flipthem}
\bibfield{author}{\bibinfo{person}{Aron Laszka}, \bibinfo{person}{Gabor Horvath}, \bibinfo{person}{Mark Felegyhazi}, {and} \bibinfo{person}{Levente Butty{\'a}n}.} \bibinfo{year}{2014}\natexlab{}.
\newblock \showarticletitle{FlipThem: Modeling targeted attacks with FlipIt for multiple resources}. In \bibinfo{booktitle}{\emph{International Conference on Decision and Game Theory for Security}}. Springer, \bibinfo{pages}{175--194}.
\newblock


\bibitem[Lei et~al\mbox{.}(2017)]%
        {lei2017optimal}
\bibfield{author}{\bibinfo{person}{Cheng Lei}, \bibinfo{person}{Duo-He Ma}, {and} \bibinfo{person}{Hong-Qi Zhang}.} \bibinfo{year}{2017}\natexlab{}.
\newblock \showarticletitle{Optimal strategy selection for moving target defense based on Markov game}.
\newblock \bibinfo{journal}{\emph{IEEE Access}}  \bibinfo{volume}{5} (\bibinfo{year}{2017}), \bibinfo{pages}{156--169}.
\newblock


\bibitem[Lei et~al\mbox{.}(2018)]%
        {lei2018incomplete}
\bibfield{author}{\bibinfo{person}{Cheng Lei}, \bibinfo{person}{Hong-Qi Zhang}, \bibinfo{person}{Li-Ming Wan}, \bibinfo{person}{Lu Liu}, {and} \bibinfo{person}{Duo-he Ma}.} \bibinfo{year}{2018}\natexlab{}.
\newblock \showarticletitle{Incomplete information Markov game theoretic approach to strategy generation for moving target defense}.
\newblock \bibinfo{journal}{\emph{Computer Communications}}  \bibinfo{volume}{116} (\bibinfo{year}{2018}), \bibinfo{pages}{184--199}.
\newblock


\bibitem[Li et~al\mbox{.}(2020)]%
        {li2020glide}
\bibfield{author}{\bibinfo{person}{Qianmu Li}, \bibinfo{person}{Jun Hou}, \bibinfo{person}{Shunmei Meng}, {and} \bibinfo{person}{Huaqiu Long}.} \bibinfo{year}{2020}\natexlab{}.
\newblock \showarticletitle{GLIDE: a game theory and data-driven mimicking linkage intrusion detection for edge computing networks}.
\newblock \bibinfo{journal}{\emph{Complexity}}  \bibinfo{volume}{2020} (\bibinfo{year}{2020}), \bibinfo{pages}{1--18}.
\newblock


\bibitem[Li et~al\mbox{.}(2017)]%
        {li2017differential}
\bibfield{author}{\bibinfo{person}{Zhi Li}, \bibinfo{person}{Haitao Xu}, {and} \bibinfo{person}{Yanzhu Liu}.} \bibinfo{year}{2017}\natexlab{}.
\newblock \showarticletitle{A differential game model of intrusion detection system in cloud computing}.
\newblock \bibinfo{journal}{\emph{International Journal of Distributed Sensor Networks}} \bibinfo{volume}{13}, \bibinfo{number}{1} (\bibinfo{year}{2017}), \bibinfo{pages}{1550147716687995}.
\newblock


\bibitem[Liu et~al\mbox{.}(2006)]%
        {liu2006bayesian}
\bibfield{author}{\bibinfo{person}{Yu Liu}, \bibinfo{person}{Cristina Comaniciu}, {and} \bibinfo{person}{Hong Man}.} \bibinfo{year}{2006}\natexlab{}.
\newblock \showarticletitle{A Bayesian game approach for intrusion detection in wireless ad hoc networks}. In \bibinfo{booktitle}{\emph{Proceeding from the 2006 workshop on Game theory for communications and networks}}. \bibinfo{pages}{4--es}.
\newblock


\bibitem[Maleki et~al\mbox{.}(2016)]%
        {maleki2016markov}
\bibfield{author}{\bibinfo{person}{Hoda Maleki}, \bibinfo{person}{Saeed Valizadeh}, \bibinfo{person}{William Koch}, \bibinfo{person}{Azer Bestavros}, {and} \bibinfo{person}{Marten Van~Dijk}.} \bibinfo{year}{2016}\natexlab{}.
\newblock \showarticletitle{Markov modeling of moving target defense games}. In \bibinfo{booktitle}{\emph{Proceedings of the 2016 ACM workshop on moving target defense}}. \bibinfo{pages}{81--92}.
\newblock


\bibitem[Manshaei et~al\mbox{.}(2013)]%
        {10.1145/2480741.2480742}
\bibfield{author}{\bibinfo{person}{Mohammad~Hossein Manshaei}, \bibinfo{person}{Quanyan Zhu}, \bibinfo{person}{Tansu Alpcan}, \bibinfo{person}{Tamer Bac\c{s}ar}, {and} \bibinfo{person}{Jean-Pierre Hubaux}.} \bibinfo{year}{2013}\natexlab{}.
\newblock \showarticletitle{Game Theory Meets Network Security and Privacy}.
\newblock \bibinfo{journal}{\emph{ACM Comput. Surv.}} \bibinfo{volume}{45}, \bibinfo{number}{3} (\bibinfo{date}{jul} \bibinfo{year}{2013}).
\newblock


\bibitem[Meakins(2019)]%
        {meakins2019zero}
\bibfield{author}{\bibinfo{person}{Joss Meakins}.} \bibinfo{year}{2019}\natexlab{}.
\newblock \showarticletitle{A zero-sum game: the zero-day market in 2018}.
\newblock \bibinfo{journal}{\emph{Journal of Cyber Policy}} \bibinfo{volume}{4}, \bibinfo{number}{1} (\bibinfo{year}{2019}), \bibinfo{pages}{60--71}.
\newblock


\bibitem[Min et~al\mbox{.}(2018)]%
        {min2018defense}
\bibfield{author}{\bibinfo{person}{Minghui Min}, \bibinfo{person}{Liang Xiao}, \bibinfo{person}{Caixia Xie}, \bibinfo{person}{Mohammad Hajimirsadeghi}, {and} \bibinfo{person}{Narayan~B Mandayam}.} \bibinfo{year}{2018}\natexlab{}.
\newblock \showarticletitle{Defense against advanced persistent threats in dynamic cloud storage: A colonel blotto game approach}.
\newblock \bibinfo{journal}{\emph{IEEE Internet of Things Journal}} \bibinfo{volume}{5}, \bibinfo{number}{6} (\bibinfo{year}{2018}), \bibinfo{pages}{4250--4261}.
\newblock


\bibitem[Nagurney and Shukla(2017)]%
        {nagurney2017multifirm}
\bibfield{author}{\bibinfo{person}{Anna Nagurney} {and} \bibinfo{person}{Shivani Shukla}.} \bibinfo{year}{2017}\natexlab{}.
\newblock \showarticletitle{Multifirm models of cybersecurity investment competition vs. cooperation and network vulnerability}.
\newblock \bibinfo{journal}{\emph{European Journal of Operational Research}} \bibinfo{volume}{260}, \bibinfo{number}{2} (\bibinfo{year}{2017}), \bibinfo{pages}{588--600}.
\newblock


\bibitem[Niazi and Faheem(2019)]%
        {niazi2019bayesian}
\bibfield{author}{\bibinfo{person}{Rumaisa~Aimen Niazi} {and} \bibinfo{person}{Yasir Faheem}.} \bibinfo{year}{2019}\natexlab{}.
\newblock \showarticletitle{A bayesian game-theoretic intrusion detection system for hypervisor-based software defined networks in smart grids}.
\newblock \bibinfo{journal}{\emph{IEEE Access}}  \bibinfo{volume}{7} (\bibinfo{year}{2019}), \bibinfo{pages}{88656--88672}.
\newblock


\bibitem[Panda et~al\mbox{.}(2022)]%
        {panda2022honeycar}
\bibfield{author}{\bibinfo{person}{Sakshyam Panda}, \bibinfo{person}{Stefan Rass}, \bibinfo{person}{Sotiris Moschoyiannis}, \bibinfo{person}{Kaitai Liang}, \bibinfo{person}{George Loukas}, {and} \bibinfo{person}{Emmanouil Panaousis}.} \bibinfo{year}{2022}\natexlab{}.
\newblock \showarticletitle{HoneyCar: a framework to configure honeypot vulnerabilities on the internet of vehicles}.
\newblock \bibinfo{journal}{\emph{IEEE Access}}  \bibinfo{volume}{10} (\bibinfo{year}{2022}), \bibinfo{pages}{104671--104685}.
\newblock


\bibitem[Papadimitriou and Piliouras(2016)]%
        {Papadimitriou2016}
\bibfield{author}{\bibinfo{person}{Christos Papadimitriou} {and} \bibinfo{person}{Georgios Piliouras}.} \bibinfo{year}{2016}\natexlab{}.
\newblock \showarticletitle{{From Nash Equilibria to Chain Recurrent Sets: Solution Concepts and Topology}}. In \bibinfo{booktitle}{\emph{ACM Conference on Innovations in Theoretical Computer Science}}. \bibinfo{pages}{227--235}.
\newblock


\bibitem[Pawlick et~al\mbox{.}(2019)]%
        {pawlick2019game}
\bibfield{author}{\bibinfo{person}{Jeffrey Pawlick}, \bibinfo{person}{Edward Colbert}, {and} \bibinfo{person}{Quanyan Zhu}.} \bibinfo{year}{2019}\natexlab{}.
\newblock \showarticletitle{A game-theoretic taxonomy and survey of defensive deception for cybersecurity and privacy}.
\newblock \bibinfo{journal}{\emph{ACM Computing Surveys (CSUR)}} \bibinfo{volume}{52}, \bibinfo{number}{4} (\bibinfo{year}{2019}), \bibinfo{pages}{1--28}.
\newblock


\bibitem[Pawlick et~al\mbox{.}(2015)]%
        {pawlick2015flip}
\bibfield{author}{\bibinfo{person}{Jeffrey Pawlick}, \bibinfo{person}{Sadegh Farhang}, {and} \bibinfo{person}{Quanyan Zhu}.} \bibinfo{year}{2015}\natexlab{}.
\newblock \showarticletitle{Flip the cloud: Cyber-physical signaling games in the presence of advanced persistent threats}. In \bibinfo{booktitle}{\emph{International Conference on Decision and Game Theory for Security}}. Springer, \bibinfo{pages}{289--308}.
\newblock


\bibitem[Pedregal(2004)]%
        {pedregal2004introduction}
\bibfield{author}{\bibinfo{person}{Pablo Pedregal}.} \bibinfo{year}{2004}\natexlab{}.
\newblock \bibinfo{booktitle}{\emph{Introduction to optimization}}. Vol.~\bibinfo{volume}{46}.
\newblock \bibinfo{publisher}{Springer}.
\newblock


\bibitem[P{\'\i}bil et~al\mbox{.}(2012)]%
        {pibil2012game}
\bibfield{author}{\bibinfo{person}{Radek P{\'\i}bil}, \bibinfo{person}{Viliam Lis{\`y}}, \bibinfo{person}{Christopher Kiekintveld}, \bibinfo{person}{Branislav Bo{\v{s}}ansk{\`y}}, {and} \bibinfo{person}{Michal P{\v{e}}chou{\v{c}}ek}.} \bibinfo{year}{2012}\natexlab{}.
\newblock \showarticletitle{Game theoretic model of strategic honeypot selection in computer networks}. In \bibinfo{booktitle}{\emph{Decision and Game Theory for Security: Third International Conference, GameSec 2012}}. Springer.
\newblock


\bibitem[Pradelski and Young(2012)]%
        {Pradelski2012}
\bibfield{author}{\bibinfo{person}{Bary Pradelski} {and} \bibinfo{person}{H.~Peyton Young}.} \bibinfo{year}{2012}\natexlab{}.
\newblock \showarticletitle{{Learning Efficient Nash Equilibria in Distributed Systems}}.
\newblock \bibinfo{journal}{\emph{Games and Economic Behavior}} \bibinfo{volume}{75}, \bibinfo{number}{2} (\bibinfo{year}{2012}), \bibinfo{pages}{882--897}.
\newblock


\bibitem[Prakash and Wellman(2015)]%
        {prakash2015empirical}
\bibfield{author}{\bibinfo{person}{Achintya Prakash} {and} \bibinfo{person}{Michael~P Wellman}.} \bibinfo{year}{2015}\natexlab{}.
\newblock \showarticletitle{Empirical game-theoretic analysis for moving target defense}. In \bibinfo{booktitle}{\emph{Proceedings of the second ACM workshop on moving target defense}}. \bibinfo{pages}{57--65}.
\newblock


\bibitem[Rass et~al\mbox{.}(2019)]%
        {rass2019cut}
\bibfield{author}{\bibinfo{person}{Stefan Rass}, \bibinfo{person}{Sandra K{\"o}nig}, {and} \bibinfo{person}{Emmanouil Panaousis}.} \bibinfo{year}{2019}\natexlab{}.
\newblock \showarticletitle{Cut-the-rope: A game of stealthy intrusion}. In \bibinfo{booktitle}{\emph{Decision and Game Theory for Security: 10th International Conference, GameSec 2019, Stockholm, Sweden, October 30--November 1, 2019, Proceedings}}. Springer, \bibinfo{pages}{404--416}.
\newblock


\bibitem[Rass et~al\mbox{.}(2017)]%
        {rass2017defending}
\bibfield{author}{\bibinfo{person}{Stefan Rass}, \bibinfo{person}{Sandra K{\"o}nig}, {and} \bibinfo{person}{Stefan Schauer}.} \bibinfo{year}{2017}\natexlab{}.
\newblock \showarticletitle{Defending against advanced persistent threats using game-theory}.
\newblock \bibinfo{journal}{\emph{PloS one}} \bibinfo{volume}{12}, \bibinfo{number}{1} (\bibinfo{year}{2017}), \bibinfo{pages}{e0168675}.
\newblock


\bibitem[Sengupta et~al\mbox{.}(2018)]%
        {sengupta2018moving}
\bibfield{author}{\bibinfo{person}{Sailik Sengupta}, \bibinfo{person}{Ankur Chowdhary}, \bibinfo{person}{Dijiang Huang}, {and} \bibinfo{person}{Subbarao Kambhampati}.} \bibinfo{year}{2018}\natexlab{}.
\newblock \showarticletitle{Moving target defense for the placement of intrusion detection systems in the cloud}. In \bibinfo{booktitle}{\emph{International Conference on Decision and Game Theory for Security}}. Springer, \bibinfo{pages}{326--345}.
\newblock


\bibitem[Sengupta et~al\mbox{.}(2020)]%
        {sengupta2020survey}
\bibfield{author}{\bibinfo{person}{Sailik Sengupta}, \bibinfo{person}{Ankur Chowdhary}, \bibinfo{person}{Abdulhakim Sabur}, \bibinfo{person}{Adel Alshamrani}, \bibinfo{person}{Dijiang Huang}, {and} \bibinfo{person}{Subbarao Kambhampati}.} \bibinfo{year}{2020}\natexlab{}.
\newblock \showarticletitle{A survey of moving target defenses for network security}.
\newblock \bibinfo{journal}{\emph{IEEE Communications Surveys \& Tutorials}} \bibinfo{volume}{22}, \bibinfo{number}{3} (\bibinfo{year}{2020}), \bibinfo{pages}{1909--1941}.
\newblock


\bibitem[Sengupta et~al\mbox{.}(2017)]%
        {sengupta2017game}
\bibfield{author}{\bibinfo{person}{Sailik Sengupta}, \bibinfo{person}{Satya~Gautam Vadlamudi}, \bibinfo{person}{Subbarao Kambhampati}, \bibinfo{person}{Adam Doup{\'e}}, \bibinfo{person}{Ziming Zhao}, \bibinfo{person}{Marthony Taguinod}, {and} \bibinfo{person}{Gail-Joon Ahn}.} \bibinfo{year}{2017}\natexlab{}.
\newblock \showarticletitle{A Game Theoretic Approach to Strategy Generation for Moving Target Defense in Web Applications.}. In \bibinfo{booktitle}{\emph{AAMAS}}, Vol.~\bibinfo{volume}{1}. \bibinfo{pages}{178--186}.
\newblock


\bibitem[Shi et~al\mbox{.}(2021)]%
        {shi2021research}
\bibfield{author}{\bibinfo{person}{Leyi Shi}, \bibinfo{person}{Xiran Wang}, {and} \bibinfo{person}{Huiwen Hou}.} \bibinfo{year}{2021}\natexlab{}.
\newblock \showarticletitle{Research on optimization of array honeypot defense strategies based on evolutionary game theory}.
\newblock \bibinfo{journal}{\emph{Mathematics}} \bibinfo{volume}{9}, \bibinfo{number}{8} (\bibinfo{year}{2021}), \bibinfo{pages}{805}.
\newblock


\bibitem[Subba et~al\mbox{.}(2018)]%
        {subba2018game}
\bibfield{author}{\bibinfo{person}{Basant Subba}, \bibinfo{person}{Santosh Biswas}, {and} \bibinfo{person}{Sushanta Karmakar}.} \bibinfo{year}{2018}\natexlab{}.
\newblock \showarticletitle{A game theory based multi layered intrusion detection framework for VANET}.
\newblock \bibinfo{journal}{\emph{Future Generation Computer Systems}}  \bibinfo{volume}{82} (\bibinfo{year}{2018}), \bibinfo{pages}{12--28}.
\newblock


\bibitem[Tan et~al\mbox{.}(2019)]%
        {tan2019optimal}
\bibfield{author}{\bibinfo{person}{Jing-lei Tan}, \bibinfo{person}{Cheng Lei}, \bibinfo{person}{Hong-qi Zhang}, {and} \bibinfo{person}{Yu-qiao Cheng}.} \bibinfo{year}{2019}\natexlab{}.
\newblock \showarticletitle{Optimal strategy selection approach to moving target defense based on Markov robust game}.
\newblock \bibinfo{journal}{\emph{computers \& security}}  \bibinfo{volume}{85} (\bibinfo{year}{2019}), \bibinfo{pages}{63--76}.
\newblock


\bibitem[Tian et~al\mbox{.}(2021)]%
        {tian2021honeypot}
\bibfield{author}{\bibinfo{person}{Wen Tian}, \bibinfo{person}{Miao Du}, \bibinfo{person}{Xiaopeng Ji}, \bibinfo{person}{Guangjie Liu}, \bibinfo{person}{Yuewei Dai}, {and} \bibinfo{person}{Zhu Han}.} \bibinfo{year}{2021}\natexlab{}.
\newblock \showarticletitle{Honeypot detection strategy against advanced persistent threats in industrial internet of things: A prospect theoretic game}.
\newblock \bibinfo{journal}{\emph{IEEE Internet of Things Journal}} \bibinfo{volume}{8}, \bibinfo{number}{24} (\bibinfo{year}{2021}), \bibinfo{pages}{17372--17381}.
\newblock


\bibitem[Tian et~al\mbox{.}(2019)]%
        {tian2019defense}
\bibfield{author}{\bibinfo{person}{Wen Tian}, \bibinfo{person}{Xiaopeng Ji}, \bibinfo{person}{Weiwei Liu}, \bibinfo{person}{Guangjie Liu}, \bibinfo{person}{Rong Lin}, \bibinfo{person}{Jiangtao Zhai}, {and} \bibinfo{person}{Yuewei Dai}.} \bibinfo{year}{2019}\natexlab{}.
\newblock \showarticletitle{Defense strategies against network attacks in cyber-physical systems with analysis cost constraint based on honeypot game model}.
\newblock \bibinfo{journal}{\emph{Comput. Mater. Continua}} \bibinfo{volume}{60}, \bibinfo{number}{1} (\bibinfo{year}{2019}), \bibinfo{pages}{193--211}.
\newblock


\bibitem[Tosh et~al\mbox{.}(2015b)]%
        {tosh2015evolutionary}
\bibfield{author}{\bibinfo{person}{Deepak Tosh}, \bibinfo{person}{Shamik Sengupta}, \bibinfo{person}{Charles Kamhoua}, \bibinfo{person}{Kevin Kwiat}, {and} \bibinfo{person}{Andrew Martin}.} \bibinfo{year}{2015}\natexlab{b}.
\newblock \showarticletitle{An evolutionary game-theoretic framework for cyber-threat information sharing}. In \bibinfo{booktitle}{\emph{2015 IEEE International Conference on Communications (ICC)}}. IEEE, \bibinfo{pages}{7341--7346}.
\newblock


\bibitem[Tosh et~al\mbox{.}(2018)]%
        {tosh2018establishing}
\bibfield{author}{\bibinfo{person}{Deepak Tosh}, \bibinfo{person}{Shamik Sengupta}, \bibinfo{person}{Charles~A Kamhoua}, {and} \bibinfo{person}{Kevin~A Kwiat}.} \bibinfo{year}{2018}\natexlab{}.
\newblock \showarticletitle{Establishing evolutionary game models for cyber security information exchange (cybex)}.
\newblock \bibinfo{journal}{\emph{J. Comput. System Sci.}}  \bibinfo{volume}{98} (\bibinfo{year}{2018}), \bibinfo{pages}{27--52}.
\newblock


\bibitem[Tosh et~al\mbox{.}(2015a)]%
        {tosh2015cyber}
\bibfield{author}{\bibinfo{person}{Deepak~K Tosh}, \bibinfo{person}{Matthew Molloy}, \bibinfo{person}{Shamik Sengupta}, \bibinfo{person}{Charles~A Kamhoua}, {and} \bibinfo{person}{Kevin~A Kwiat}.} \bibinfo{year}{2015}\natexlab{a}.
\newblock \showarticletitle{Cyber-investment and cyber-information exchange decision modeling}. In \bibinfo{booktitle}{\emph{2015 IEEE 17th International Conference on High Performance Computing and Communications, 2015 IEEE 7th International Symposium on Cyberspace Safety and Security, and 2015 IEEE 12th International Conference on Embedded Software and Systems}}. IEEE, \bibinfo{pages}{1219--1224}.
\newblock


\bibitem[Tsvetanov and Slaria(2021)]%
        {tsvetanov2021effect}
\bibfield{author}{\bibinfo{person}{Tsvetan Tsvetanov} {and} \bibinfo{person}{Srishti Slaria}.} \bibinfo{year}{2021}\natexlab{}.
\newblock \showarticletitle{The effect of the Colonial Pipeline shutdown on gasoline prices}.
\newblock \bibinfo{journal}{\emph{Economics Letters}}  \bibinfo{volume}{209} (\bibinfo{year}{2021}), \bibinfo{pages}{110122}.
\newblock


\bibitem[Umsonst and Sandberg(2018)]%
        {umsonst2018game}
\bibfield{author}{\bibinfo{person}{David Umsonst} {and} \bibinfo{person}{Henrik Sandberg}.} \bibinfo{year}{2018}\natexlab{}.
\newblock \showarticletitle{A game-theoretic approach for choosing a detector tuning under stealthy sensor data attacks}. In \bibinfo{booktitle}{\emph{2018 IEEE Conference on Decision and Control (CDC)}}. IEEE, \bibinfo{pages}{5975--5981}.
\newblock


\bibitem[Vakilinia and Sengupta(2017)]%
        {vakilinia2017coalitional}
\bibfield{author}{\bibinfo{person}{Iman Vakilinia} {and} \bibinfo{person}{Shamik Sengupta}.} \bibinfo{year}{2017}\natexlab{}.
\newblock \showarticletitle{A coalitional game theory approach for cybersecurity information sharing}. In \bibinfo{booktitle}{\emph{MILCOM 2017-2017 IEEE Military Communications Conference (MILCOM)}}. IEEE, \bibinfo{pages}{237--242}.
\newblock


\bibitem[Vakilinia and Sengupta(2018)]%
        {vakilinia2018coalitional}
\bibfield{author}{\bibinfo{person}{Iman Vakilinia} {and} \bibinfo{person}{Shamik Sengupta}.} \bibinfo{year}{2018}\natexlab{}.
\newblock \showarticletitle{A coalitional cyber-insurance framework for a common platform}.
\newblock \bibinfo{journal}{\emph{IEEE Transactions on Information Forensics and Security}} \bibinfo{volume}{14}, \bibinfo{number}{6} (\bibinfo{year}{2018}), \bibinfo{pages}{1526--1538}.
\newblock


\bibitem[Vakilinia and Sengupta(2019)]%
        {vakilinia2019fair}
\bibfield{author}{\bibinfo{person}{Iman Vakilinia} {and} \bibinfo{person}{Shamik Sengupta}.} \bibinfo{year}{2019}\natexlab{}.
\newblock \showarticletitle{Fair and private rewarding in a coalitional game of cybersecurity information sharing}.
\newblock \bibinfo{journal}{\emph{IET Information Security}} \bibinfo{volume}{13}, \bibinfo{number}{6} (\bibinfo{year}{2019}), \bibinfo{pages}{530--540}.
\newblock


\bibitem[Van~Dijk et~al\mbox{.}(2013)]%
        {van2013flipit}
\bibfield{author}{\bibinfo{person}{Marten Van~Dijk}, \bibinfo{person}{Ari Juels}, \bibinfo{person}{Alina Oprea}, {and} \bibinfo{person}{Ronald~L Rivest}.} \bibinfo{year}{2013}\natexlab{}.
\newblock \showarticletitle{FlipIt: The game of “stealthy takeover”}.
\newblock \bibinfo{journal}{\emph{Journal of Cryptology}} \bibinfo{volume}{26}, \bibinfo{number}{4} (\bibinfo{year}{2013}), \bibinfo{pages}{655--713}.
\newblock


\bibitem[Wang et~al\mbox{.}(2017)]%
        {wang2017strategic}
\bibfield{author}{\bibinfo{person}{Kun Wang}, \bibinfo{person}{Miao Du}, \bibinfo{person}{Sabita Maharjan}, {and} \bibinfo{person}{Yanfei Sun}.} \bibinfo{year}{2017}\natexlab{}.
\newblock \showarticletitle{Strategic honeypot game model for distributed denial of service attacks in the smart grid}.
\newblock \bibinfo{journal}{\emph{IEEE Transactions on Smart Grid}} \bibinfo{volume}{8}, \bibinfo{number}{5} (\bibinfo{year}{2017}), \bibinfo{pages}{2474--2482}.
\newblock


\bibitem[Wang et~al\mbox{.}(2019)]%
        {wang2019moving}
\bibfield{author}{\bibinfo{person}{Shengling Wang}, \bibinfo{person}{Hongwei Shi}, \bibinfo{person}{Qin Hu}, \bibinfo{person}{Bin Lin}, {and} \bibinfo{person}{Xiuzhen Cheng}.} \bibinfo{year}{2019}\natexlab{}.
\newblock \showarticletitle{Moving target defense for internet of things based on the zero-determinant theory}.
\newblock \bibinfo{journal}{\emph{IEEE Internet of Things Journal}} \bibinfo{volume}{7}, \bibinfo{number}{1} (\bibinfo{year}{2019}), \bibinfo{pages}{661--668}.
\newblock


\bibitem[Wang et~al\mbox{.}(2022)]%
        {wang2022learning}
\bibfield{author}{\bibinfo{person}{Yuntao Wang}, \bibinfo{person}{Zhou Su}, \bibinfo{person}{Abderrahim Benslimane}, \bibinfo{person}{Qichao Xu}, \bibinfo{person}{Minghui Dai}, {and} \bibinfo{person}{Ruidong Li}.} \bibinfo{year}{2022}\natexlab{}.
\newblock \showarticletitle{A Learning-based Honeypot Game for Collaborative Defense in UAV Networks}. In \bibinfo{booktitle}{\emph{GLOBECOM 2022-2022 IEEE Global Communications Conference}}. IEEE, \bibinfo{pages}{3521--3526}.
\newblock


\bibitem[Xiao et~al\mbox{.}(2018)]%
        {xiao2018attacker}
\bibfield{author}{\bibinfo{person}{Liang Xiao}, \bibinfo{person}{Dongjin Xu}, \bibinfo{person}{Narayan~B Mandayam}, {and} \bibinfo{person}{H~Vincent Poor}.} \bibinfo{year}{2018}\natexlab{}.
\newblock \showarticletitle{Attacker-centric view of a detection game against advanced persistent threats}.
\newblock \bibinfo{journal}{\emph{IEEE transactions on mobile computing}} \bibinfo{volume}{17}, \bibinfo{number}{11} (\bibinfo{year}{2018}), \bibinfo{pages}{2512--2523}.
\newblock


\bibitem[Xiao et~al\mbox{.}(2017)]%
        {xiao2017cloud}
\bibfield{author}{\bibinfo{person}{Liang Xiao}, \bibinfo{person}{Dongjin Xu}, \bibinfo{person}{Caixia Xie}, \bibinfo{person}{Narayan~B Mandayam}, {and} \bibinfo{person}{H~Vincent Poor}.} \bibinfo{year}{2017}\natexlab{}.
\newblock \showarticletitle{Cloud storage defense against advanced persistent threats: A prospect theoretic study}.
\newblock \bibinfo{journal}{\emph{IEEE Journal on Selected Areas in Communications}} \bibinfo{volume}{35}, \bibinfo{number}{3} (\bibinfo{year}{2017}), \bibinfo{pages}{534--544}.
\newblock


\bibitem[Xu(2021)]%
        {XuSciSec2021SARR}
\bibfield{author}{\bibinfo{person}{Shouhuai Xu}.} \bibinfo{year}{2021}\natexlab{}.
\newblock \showarticletitle{SARR: A Cybersecurity Metrics and Quantification Framework}. In \bibinfo{booktitle}{\emph{Third International Conference on Science of Cyber Security (SciSec'2021)}}. \bibinfo{pages}{3--17}.
\newblock


\bibitem[Yang et~al\mbox{.}(2018a)]%
        {yang2018risk}
\bibfield{author}{\bibinfo{person}{Lu-Xing Yang}, \bibinfo{person}{Pengdeng Li}, \bibinfo{person}{Xiaofan Yang}, {and} \bibinfo{person}{Yuan~Yan Tang}.} \bibinfo{year}{2018}\natexlab{a}.
\newblock \showarticletitle{A risk management approach to defending against the advanced persistent threat}.
\newblock \bibinfo{journal}{\emph{IEEE Transactions on Dependable and Secure Computing}} \bibinfo{volume}{17}, \bibinfo{number}{6} (\bibinfo{year}{2018}), \bibinfo{pages}{1163--1172}.
\newblock


\bibitem[Yang et~al\mbox{.}(2018b)]%
        {yang2018effective}
\bibfield{author}{\bibinfo{person}{Lu-Xing Yang}, \bibinfo{person}{Pengdeng Li}, \bibinfo{person}{Yushu Zhang}, \bibinfo{person}{Xiaofan Yang}, \bibinfo{person}{Yong Xiang}, {and} \bibinfo{person}{Wanlei Zhou}.} \bibinfo{year}{2018}\natexlab{b}.
\newblock \showarticletitle{Effective repair strategy against advanced persistent threat: A differential game approach}.
\newblock \bibinfo{journal}{\emph{IEEE Transactions on Information Forensics and Security}} \bibinfo{volume}{14}, \bibinfo{number}{7} (\bibinfo{year}{2018}), \bibinfo{pages}{1713--1728}.
\newblock


\bibitem[Zhang et~al\mbox{.}(2014)]%
        {zhang2014stealthy}
\bibfield{author}{\bibinfo{person}{Ming Zhang}, \bibinfo{person}{Zizhan Zheng}, {and} \bibinfo{person}{Ness~B Shroff}.} \bibinfo{year}{2014}\natexlab{}.
\newblock \showarticletitle{Stealthy attacks and observable defenses: A game theoretic model under strict resource constraints}. In \bibinfo{booktitle}{\emph{2014 IEEE Global Conference on Signal and Information Processing (GlobalSIP)}}. IEEE, \bibinfo{pages}{813--817}.
\newblock


\bibitem[Zhang et~al\mbox{.}(2015)]%
        {zhang2015game}
\bibfield{author}{\bibinfo{person}{Ming Zhang}, \bibinfo{person}{Zizhan Zheng}, {and} \bibinfo{person}{Ness~B Shroff}.} \bibinfo{year}{2015}\natexlab{}.
\newblock \showarticletitle{A game theoretic model for defending against stealthy attacks with limited resources}. In \bibinfo{booktitle}{\emph{Decision and Game Theory for Security: 6th International Conference, GameSec 2015, London, UK, November 4-5, 2015, Proceedings 6}}. Springer, \bibinfo{pages}{93--112}.
\newblock


\bibitem[Zhu et~al\mbox{.}(2021)]%
        {zhu2021survey}
\bibfield{author}{\bibinfo{person}{Mu Zhu}, \bibinfo{person}{Ahmed~H Anwar}, \bibinfo{person}{Zelin Wan}, \bibinfo{person}{Jin-Hee Cho}, \bibinfo{person}{Charles~A Kamhoua}, {and} \bibinfo{person}{Munindar~P Singh}.} \bibinfo{year}{2021}\natexlab{}.
\newblock \showarticletitle{A survey of defensive deception: Approaches using game theory and machine learning}.
\newblock \bibinfo{journal}{\emph{IEEE Communications Surveys \& Tutorials}} \bibinfo{volume}{23}, \bibinfo{number}{4} (\bibinfo{year}{2021}), \bibinfo{pages}{2460--2493}.
\newblock


\bibitem[Zhu and Ba{\c{s}}ar(2011)]%
        {zhu2011robust}
\bibfield{author}{\bibinfo{person}{Quanyan Zhu} {and} \bibinfo{person}{Tamer Ba{\c{s}}ar}.} \bibinfo{year}{2011}\natexlab{}.
\newblock \showarticletitle{Robust and resilient control design for cyber-physical systems with an application to power systems}. In \bibinfo{booktitle}{\emph{2011 50th IEEE Conference on Decision and Control and European Control Conference}}. IEEE, \bibinfo{pages}{4066--4071}.
\newblock


\bibitem[Zhu and Ba{\c{s}}ar(2013)]%
        {zhu2013game}
\bibfield{author}{\bibinfo{person}{Quanyan Zhu} {and} \bibinfo{person}{Tamer Ba{\c{s}}ar}.} \bibinfo{year}{2013}\natexlab{}.
\newblock \showarticletitle{Game-theoretic approach to feedback-driven multi-stage moving target defense}. In \bibinfo{booktitle}{\emph{International conference on decision and game theory for security}}. Springer, \bibinfo{pages}{246--263}.
\newblock


\bibitem[Zhu and Rass(2018)]%
        {zhu2018multi}
\bibfield{author}{\bibinfo{person}{Quanyan Zhu} {and} \bibinfo{person}{Stefan Rass}.} \bibinfo{year}{2018}\natexlab{}.
\newblock \showarticletitle{On multi-phase and multi-stage game-theoretic modeling of advanced persistent threats}.
\newblock \bibinfo{journal}{\emph{IEEE Access}}  \bibinfo{volume}{6} (\bibinfo{year}{2018}), \bibinfo{pages}{13958--13971}.
\newblock


\bibitem[Zibak and Simpson(2019)]%
        {zibak2019cyber}
\bibfield{author}{\bibinfo{person}{Adam Zibak} {and} \bibinfo{person}{Andrew Simpson}.} \bibinfo{year}{2019}\natexlab{}.
\newblock \showarticletitle{Cyber threat information sharing: Perceived benefits and barriers}. In \bibinfo{booktitle}{\emph{Proceedings of the 14th international conference on availability, reliability and security}}. \bibinfo{pages}{1--9}.
\newblock


\end{thebibliography}

\section{Appendix: Key Analysis Techniques}
\label{sec:analysis techniques}

We systematize the techniques that have been used to analyze the models discussed in the main text of the paper. These techniques are highlighted in the ``Analysis Technique'' column in Table \ref{tab:assumptions}. 
We here focus on some popular techniques in the following subsections while referring to the respective publications for all undiscussed techniques.


\subsection{Dominant Strategy (DS)}
\label{sec:DS}
This technique pertains to two-player Normal form games.
To show the basic idea, recall the attack-defense game posed in Section~\ref{ssec:normal form game} where the attacker incurs a small cost to launch attacks and gets a large gain for successfully launching an attack.
Similarly, the defender incurs a large loss if their system is compromised.
Here we present new examples, where in the top game the defender can monitor their system for free, and in the bottom monitoring has a cost.

\begin{table}[!htbp]
\begin{center}
\begin{tabular}{ |c|cc| } 
 \hline
 free monitoring & attack & wait \\ 
 \hline
 monitor & 0,-1 & 0,0 \\ 
 wait & -5,5 & 0,0 \\ 
 \hline
\end{tabular}
\hfill
\begin{tabular}{ |c|cc| } 
 \hline
 monitor costs& attack & wait \\ 
 \hline
 monitor & -1,-1 & -1,0 \\ 
 wait & -5,5 & 0,0 \\ 
 \hline
\end{tabular}
\end{center}
\caption{Two attack defense games where a defender decides whether to monitor their system.
In each game, the left (right) entry in each box corresponds to the row (column) player's utility upon the realized event. \label{tab:numeric example}}
\end{table}


Intuitively, a dominant strategy can be seen for the defender using the monitor strategy because the defender is always better off or equal regardless of the attacker's action.
If the attacker decides to attack, the defender prevents a large loss of a possible breach by monitoring; if the attacker waits, the defender is indifferent between waiting and monitoring.
Formally, an action $a_i\in A_i$ is said to be a \emph{dominant strategy} if for all action profiles of other players $a_{-i}\in A_{-i}$ we have
\begin{equation}\label{eq:dominantion}
    u_i(a_i,a_{-i})\geq u_i(a'_i,a_{-i}) \textrm{ } \forall {a'_i\in A_i}.
\end{equation}
Eq.\eqref{eq:dominantion} formalizes the notion in the free monitoring attack-defense game, where strategy $a_i$ is dominant if player $i$ prefers it regardless of the other player's actions.

If a dominant strategy does exist, it is easy to see it satisfies the Nash equilibrium condition given by Eq.\eqref{eq:Nash} for player $i$.
Consequently, any Nash equilibrium must feature player $i$ playing their (strict) dominant strategy, should they have one.
However, dominant strategies may not exist in Normal Form games, which we can see by example on the right-hand side of Table~\ref{tab:numeric example}.
Specifically, because the defender incurs a monitoring cost, if the attacker waits, the defender also prefers to wait, since monitoring their system will waste resources.
Thus, in that example, monitor is no longer a dominant strategy, and the defender must seek some form of mixture between their two actions.

These positive and negative examples of dominant strategies represent a lesson in cybersecurity.
If defensive actions are free (or very cheap), as may be the case in moving target defense or low-interaction honeypots, then game theory can be used to justify the use of defensive mechanisms, even if the attacker does not strike frequently.
However, if the defensive action is costly, for example high-interaction honeypots, then the defender must consider how frequently to take action more carefully.
As such, dominant strategies have been used in the analysis of honeypot deployment scenarios\cite{wang2017strategic,tian2019defense}.

\subsection{Linear Complementarity Programming (LCP)} 
\label{sec:LCP}

This technique also pertains to two-player Normal Form games.
In the above, we have established that dominant strategies are a powerful tool for cyber decision-making when they exist, the natural next question is what should be done when they do not exist?
This and subsequent analysis techniques deal with this situation, where both players must react to each other's preferences and cannot rely on a single action that dominates in all possible scenarios.

%
%

The broadest answer to the preceding question is the Nash equilibrium, whose strategies represent a point in the action space that no player can profitably and unilaterally escape.
Generally, the Nash equilibrium of two-player games with arbitrary numbers of actions can be solved by formulating it as an LCP \cite{eaves1971linear}.
%
In terms of cybersecurity, this computation of the Nash equilibrium illuminates its weaknesses in application.
The most counterintuitive aspect of the calculation is that often each player does not even consider their own utilities, instead computing their action entirely based on their opponent's utility function.
In cybersecurity, there is often a high degree of uncertainty as regards to who the adversary is or what their capabilities are.
This reality is in stark contrast to the Nash computation, which relies heavily on knowing the structure of the game and how opposing players evaluate all possible situations.

\subsection{Linear Programming (LP)}
\label{sec:LP}
Consider the Stackelberg equilibrium.
Recall that Stackelberg game is defined on Normal form games just as Nash is, but additionally requires a strict order in which players make decisions.
For example, suppose that the defender in Table~\ref{tab:numeric example} must go first and that the attacker can observe their action and best respond accordingly.
Then, we can consider the defender the leader who seeks a mixed strategy $\sigma_L\in\Delta(A_L)$ to maximize their utility at Stackelberg equilibrium, formalized in Eq.\eqref{eq:Stackelberg}.

An algorithmic approach to solving the Stackelberg equilibrium was first presented in \cite{conitzer2006computing}, using $|A_F|$ linear programs.
The key intuition to compute the equilibrium strategy $\sigma^*$ is that the leader can consider each of the follower's actions individually, with the condition that the leader must constrain themselves such that a rational follower would respond to them with the said action.
This approach is effective because, due to the structure of Stackelberg games, the follower's optimal action will always be a pure strategy, greatly reducing the search space.
For a specific follower action $a_F$, this can be solved using a linear program as follows:
\begin{equation}\label{eq:Stackelberg LP}
\begin{aligned}
\max &\sum_{a_L} \sigma_L(a_L) u_L(a_L,a_F)\\\textrm{s.t. }
& \sigma_L\in\Delta(A_L) \\
&\sum_{a_L} \sigma_L(a_L) u_F(a_L,a_F)\geq \sum_{a_L} \sigma_L(a_L) u_F(a_L,a'_F) \\
&\qquad \forall a'_F\in A_F.
\end{aligned}
\end{equation}

In the above optimization problem, the objective function is simply the expected value of the leader's utility function given their strategy mix $\sigma_L$.
The first constraint ensures $\sigma_L$ is a valid probability, and the second constraint, known as {\em complementary slackness}, ensures that the pre\-selected follower action $a_F$ is an optimal best response to the leader's action.
To find the equilibrium action profile $\sigma^*_L$, iterate over each follower action $a_F$ and solve the linear program \eqref{eq:Stackelberg LP}.
Once this computation is done, the Stackelberg equilibrium action $\sigma^*_L$ is simply the $\sigma_L$ corresponding to the maximum expected leader's utility.

\subsection{Dynamic Programming (DP)}
\label{sec:DP}

This technique is also known as {\em backward induction}.
It applies to multi-stage games where decisions early in the game impact those later in the game.
The key intuition of this process is that to solve the best action at any time $t$, it is necessary to consider the utility at time $t$, $U^t_i(a^t)$, and the optimal utility incurred at all future stages, given the impact that the action $a^t$ will have on the system.

The main challenge in solving this problem is calculating the optimal utility at all future steps,  especially when the problem has many steps.
To address this, the key insight of the DP algorithm is to solve the problem starting with the final stage, which is simply optimizing the final stage utility $u^T_i$.
Then, the second to last stage can be solved as we know both the utility at that stage $u^{T-1}_i$, and the optimal solution to the future stage as well.
This process continues until the first stage is reached and an optimal policy for each stage has been computed.
The DP technique has been used to study multistage cybersecurity games,
especially APT situations with homogeneous game models between stages \cite{huang2018analysis,huang2019adaptive,huang2020dynamic}.

\subsection{Q-Learning}
\label{sec:Q learning}
Due to the complex nature of cybersecurity, known analysis techniques often cannot provide disciplined answers.
In these situations, Reinforcement Learning can be utilized in the absence of analytical techniques.
The key advantage of Reinforcement Learning over analytical techniques is that it assumes little knowledge of the structure or parameters of the problem, which often rings true in cybersecurity applications.
Reinforcement Learning techniques such as Q-learning seek to learn optimal behaviors in an environment a player initially knows little about.
The key elements of Q learning include one or more decision-makers, an environment with which the players interact, observations from the said environment, and a reward function $r$ that rewards the players for achieving their objectives. 
Q learning has been used to learn where to place IDS systems \cite{agah2004intrusion,chowdhary2018markov} and how an IDS should respond to an attacker \cite{alpcan2006intrusion}.

In the simple case, we can regard the environment as a discrete set $S$, and the player observes which state $s\in S$ they are in.
The objective of Q learning is then to develop a function $Q:S\times A\rightarrow \mathbb{R}$ that predicts the future reward for each action given the current state.
To develop this function, the Q-learning algorithm repeatedly simulates the environment and iteratively updates the Q-function.

This has the effect that rewards, which in practical scenarios are rare, get spread around the state space in the $Q$ function.
In this way, players gain an understanding of what to do in all states, even if the reward function provides little and infrequent feedback.
One issue with this approach is that it requires maintaining a table of size $|A|\times |S|$, which rapidly becomes unwieldy; however, modern Deep Learning techniques have seen some success in handling this.

\end{document}